\documentclass[english,aps,prb,amsmath,amssymb,twocolumn,letterpaper,showpacs,superscriptaddress]{revtex4-2}

\usepackage{mathtools} 

\usepackage[normalem]{ulem} 

\usepackage{comment}
\usepackage{amsmath}
\usepackage{physics}
\usepackage{amssymb}
\usepackage{pifont}
\usepackage{bbm}
\usepackage{graphicx,bm}
\usepackage{makecell,multirow}
\usepackage{babel}
\usepackage{amstext}
\usepackage{esint}
\usepackage{cases}
\usepackage{makecell}
\usepackage{lineno}
\usepackage[caption=false]{subfig}
\usepackage{verbatim}
\usepackage[unicode=true,pdfusetitle,
bookmarks=true,bookmarksnumbered=false,bookmarksopen=false, breaklinks=false,pdfborder={0 0 1},backref=false,colorlinks=false]
 {hyperref}
 \hypersetup{
    colorlinks=true,
    linkcolor=red,
    citecolor=blue,
    filecolor=magenta,      
    urlcolor=blue
    }
\usepackage{pdfpages}	
\usepackage{pgffor}		
\usepackage{upgreek}
\usepackage{braket}

\newcommand{\cmark}{\ding{51}}%
\newcommand{\xmark}{\ding{55}}%

\makeatletter
\AtBeginDocument{\let\LS@rot\@undefined} 
\makeatother							 

\begin{document}

\title{Symmetry and Topology in a Non-Hermitian Kitaev chain}

\author{Ayush Raj}
\thanks{These authors contributed equally to this work.}
\affiliation{Department of Physics and Astronomy, Purdue University, West Lafayette, Indiana 47907, USA}

\author{Soham Ray}
\thanks{These authors contributed equally to this work.}
\affiliation{Department of Physics and Astronomy, Purdue University, West Lafayette, Indiana 47907, USA}

\author{Sai Satyam Samal}
\thanks{These authors contributed equally to this work.}
\affiliation{Department of Physics and Astronomy, Purdue University, West Lafayette, Indiana 47907, USA}

\begin{abstract}

We investigate the non-Hermitian Kitaev chain with non-reciprocal hopping amplitudes and asymmetric superconducting pairing. We work out the symmetry structure of the
model and show that particle-hole symmetry (PHS) is preserved throughout the entire parameter regime. As a consequence of PHS, the topological phase transition point of a finite open chain
coincides with that of the periodic (infinite) system. By explicitly constructing the zero-energy wave functions (Majorana modes), we show that Majorana modes necessarily occur as reciprocal localization pairs accumulating on opposite boundaries, whose combined probability density exhibits an exact cancellation of the non-Hermitian skin effect for the zero energy modes. Excited states, by contrast, generically display skin-effect localization, with particle and hole components accumulating at opposite ends of the system. At the level of bulk topology, we further construct a $\mathbb{Z}_2$
topological invariant in restricted parameter regimes that correctly distinguishes the topological and trivial phases. Finally, we present the topological phase
diagram of the non-Hermitian Kitaev chain across a broad range of complex parameters and delineate the associated phase boundaries.

\end{abstract}

\maketitle

\section{Introduction}

Non-Hermitian quantum mechanics has engaged physicists for many years from various research areas, e.g. mathematical physics~\cite{PhysRevLett.80.5243, Bender_2007,  10.1063/1.532860, PhysRevLett.89.270401, Mostafazadeh_2003}, complex space quantum mechanics~\cite{BENDER1993442, PhysRevD.73.085002}, and scattering theory~\cite{MOISEYEV1998212}. More recently, researchers have been interested in understanding the quantum dynamics of open quantum systems, i.e. systems coupled with an environment~\cite{PhysRevLett.108.173901,  PhysRevA.95.022117, RevModPhys.89.015001, PhysRevA.106.L011501, Rotter_2009, PhysRevLett.124.196401} for their potential application in quantum optics~\cite{10.1088/978-0-7503-6014-2, Carmichael1993} and quantum information theory~\cite{RevModPhys.88.021002, LewisSwan2019}. Moreover, non-Hermitian generalizations of lattice models provide an excellent testbed to study phenomena ubiquitous to non-Hermitian physics, such as complex spectra, the appearance of exceptional points and the non-Hermitian skin effect (NHSE) i.e., exponential localization of the eigenstates towards one edge (or boundary)~\cite{PhysRevLett.121.086803,PhysRevLett.123.170401,PhysRevB.99.201103,PhysRevB.100.054301,PhysRevLett.124.250402,PhysRevLett.125.126402,PhysRevLett.124.086801,PhysRevLett.129.070401,Zhang_2022,PhysRevX.13.021007,PhysRevLett.132.113802,xu2025excitonicskineffect,vhz9-xwf4,wztw-l8wg}.

The Hatano-Nelson model~\cite{PhysRevLett.77.570}, a generalization of tight-binding Hamiltonian with non-reciprocal hopping amplitude is an example of a system that exhibits NHSE. An unusual feature of such systems is the high sensitivity of the spectrum to the boundary conditions. This manifests itself in the fact that under periodic boundary conditions, the spectrum is complex and the system is unstable, but surprisingly in a finite open chain, the spectrum is purely real. The eigenstates, instead of being Bloch waves, localize exponentially toward one edge, giving rise to the NHSE.

In a set of seminal work~\cite{PhysRevLett.121.086803,PhysRevLett.123.246801}, it was shown that NHSE has a deep connection with topological properties of the system. It was shown that with non-reciprocal hoppings in the Su-Schrieffer-Heeger (SSH) model, the topological phase transition point in a finite open chain does not coincide with that of an infinite (or periodic) chain. The true transition point is governed by non-Bloch bulk-boundary correspondence, which is central to topological non-Hermitian systems. This discovery has inspired a slew of research at the intersection of non-Hermitian physics and topology \cite{PhysRevLett.123.170401,PhysRevB.100.045407,PhysRevLett.124.196401,Zhang_2022,PhysRevLett.125.126402,PhysRevLett.124.086801, Okuma_2023,RevModPhys.93.015005,PhysRevLett.121.026808,PhysRevLett.123.066404, PhysRevLett.129.070401, Xiao2020, PhysRevLett.124.056802, PhysRevX.13.021007, PhysRevLett.123.170401, PhysRevB.99.201103, PhysRevLett.125.226402, PhysRevLett.122.237601, PhysRevLett.125.180403, PhysRevLett.130.017201, PhysRevB.100.054301, PhysRevLett.124.250402, PhysRevB.109.024306, PhysRevLett.126.230402, Ghatak_2019, PhysRevLett.132.113802, Kawabata_2022, PhysRevB.101.195147, PhysRevX.8.041031,PhysRevB.55.1142, PhysRevX.9.041015, cpl_41_12_127302, brighi2025pairinginducedphasetransitionnonreciprocal, PhysRevB.111.174207, vhz9-xwf4, 6lvg-7qdn, jezequel2026nonhermitianeigenvaluesmisseigenstates, v5k2-1vj9}.

\begin{figure}[b!]
    \centering
    \includegraphics[width=1.0\linewidth]{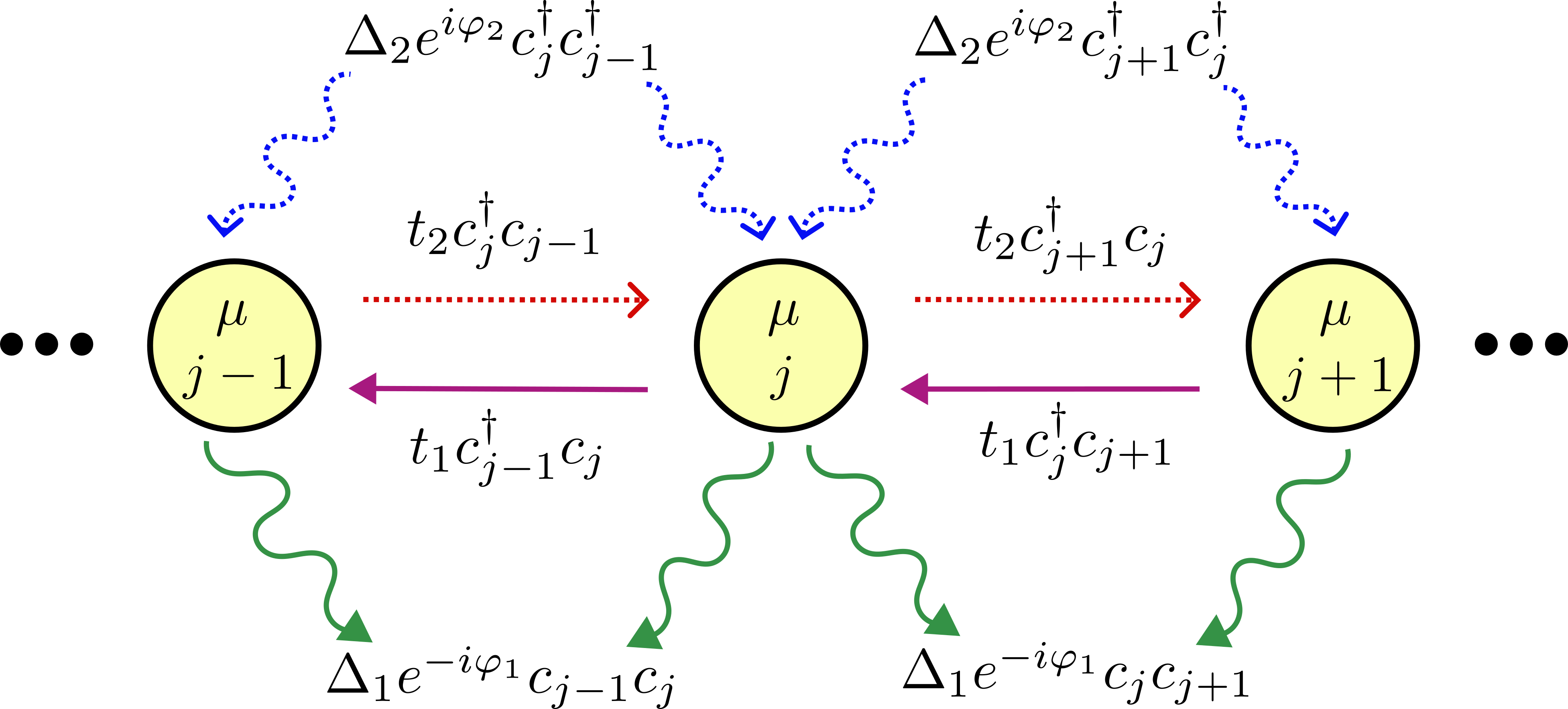}
    \caption{A caricature of a non-Hermitian Kitaev chain with non-reciprocal left and right hopping amplitudes, $t_{1}$ and $ t_{2}$, and asymmetric superconducting pairing terms, $\Delta_{1}e^{-i\varphi_1}$ and $\Delta_{2}e^{i\varphi_2}$. The right hopping is represented by a dotted arrow and the left hopping is represented by a solid arrow. Similarly, asymmetry in the superconducting pairing terms is represented by solid and dotted wave-like arrows. Each site is labeled by an integer $j$ and has a finite on-site (chemical) potential $\mu$. }
    \label{fig:non_herm_schematic}
\end{figure}

This naturally raises the question of whether such behavior is universal in one-dimensional systems. By considering a non-Hermitian Kitaev chain~\cite{Kitaev_2001,Alicea_2012,PhysRevA.94.022119,Li_2023,Yan_2023,PhysRevResearch.5.L022046,https://doi.org/10.1002/pssb.202200448,PhysRevB.111.205432,Ardonne_2025, q212-tytg} with non-reciprocal nearest neighbor hopping amplitudes ($t_{1}$ and $t_{2}$) and asymmetric superconducting pairing terms ($\Delta_{1}e^{-i\varphi_{1}}$ and $\Delta_{2}e^{i\varphi_{2}}$), see Fig.~\ref{fig:non_herm_schematic}, we show that this is not generally the case. For a non-Hermitian Kitaev chain, cf. Fig.~\ref{fig:non_herm_schematic}, the topological phase transition points for finite open chain coincides with that of an infinite (or periodic) chain. By performing a symmetry analysis, we show that particle-hole symmetry (PHS) always exists across the entire parameter regime and thereby protects the Majorana modes from NHSE. However, we show that particle and hole bulk modes exhibit an equal and opposite skin effect, which is also a consequence of PHS. Finally, we present the phase diagram for the non-Hermitian Kitaev chain in the complex chemical potential ($\mu$) plane.

This article is organized as follows. In section~\ref{sec:herm_kitaev}, we briefly review the topological phase transition in the Kitaev chain. In section~\ref{sec:non_herm_kitaev}, we introduce non-reciprocal hopping amplitudes and asymmetric superconducting pairing terms to construct a non-Hermitian Kitaev chain. Next, in section~\ref{subsec:finite_non_herm}, we numerically obtain the band structure, ground state and first excited state for a finite chain, and compare with infinite chain. In section~\ref{subsec:gbz}, we justify the robustness of Majorana zero modes to non-Hermiticity by considering the generalized Brillouin zone. In section~\ref{subsec:sym_non_herm_kitaev}, we study the symmetry structure of the non-Hermitian Kitaev chain in the complex parameter space. Finally, in section~\ref{subsec:topological_inv}, we present topological phase diagrams with different parameter choices.

\section{Kitaev chain}\label{sec:herm_kitaev}

The one-dimensional (1D) Kitaev chain exhibits topological phase transition with two distinct phases in the parameter space. It is a tight-binding chain with $p-$wave superconducting pairing terms and hosts Majorana edge states in the topological phase~\cite{Kitaev_2001}. We begin with a brief review of the 1D (Hermitian) Kitaev chain~\cite{Alicea_2012} before investigating its non-Hermitian extension. The Hamiltonian with $L$ lattice points, written in terms of particle creation ($c_{j}^{\dagger}$) and annihilation ($c_{j}$) operators on the $j^{\text{th}}$ site is given as follows,     
\begin{equation}\label{eq:herm_kitaev}
    H = -\mu \sum_{j=1}^{L} c_j^{\dagger} c_j - \frac{1}{2}\sum_{j=1}^{L} \left ( t c_j^{\dagger}c_{j+1} + \Delta e^{i\varphi} c_j c_{j+1} + h.c. \right ),
\end{equation}
where $t$ is the hopping amplitude, $\Delta$ is the superconducting gap, $\varphi$ is the superconducting phase and $\mu$ is the chemical potential. This system exhibits a topological phase transition~\cite{10.1093/acprof:oso/9780199227259.001.0001} as we tune the parameters ($t$ or $\mu$) of the Hamiltonian. The topological phase is marked by the existence of Majorana zero edge modes in a finite chain~\cite{Kitaev_2001,Alicea_2012,Aghaee2025}.

In the case of an infinite ($L\to\infty$) chain, the energy spectrum can be obtained by performing a Fourier transform i.e., substituting $c_j = \frac{1}{\sqrt{L}}\sum_k e^{-ikj} d_k$ (for simplicity, the lattice constant is set to $1$). The Hamiltonian in momentum space is of the form $H=\frac{1}{2}\sum_k \mathcal{D}_k^{\dagger} \mathcal{H}_{k} \mathcal{D}_{k}$, where, 
\begin{align}\label{eq:herm_kitaev_momentum}
    \mathcal{H}_k = \begin{pmatrix}
        -t \cos k - \mu & i\Delta e^{-i\varphi} \sin k \\
        -i\Delta e^{i\varphi} \sin k & t \cos k + \mu
    \end{pmatrix}\,,
\end{align}
and $\mathcal{D}_{k}^{\dagger} = \begin{pmatrix}
    d_{k}^{\dagger} & d_{-k}
\end{pmatrix}$. Diagonalizing $\mathcal{H}_{k}$ one obtains the dispersion relation with gap-closing at $k=0,\ \pm\pi$ and $|\mu|=t$, a necessary condition for topological phase transitions~\cite{10.1093/acprof:oso/9780199227259.001.0001,Alicea_2012}.

In topological phase, ground state sub-space of the Kitaev chain, Eq.~\eqref{eq:herm_kitaev} is two-dimensional, corresponding to free dangling Majorana modes at two ends of the chain. The system remains in a topological phase whenever $|\mu|<t$  and in a trivial phase otherwise where there are no edge states. The distinction between trivial and topological phase can be compactly captured by defining a $\mathbb{Z}_2-$topological invariant. One writes the Hamiltonian, Eq.~\eqref{eq:herm_kitaev_momentum} in the form $\mathcal{H}_k = \vec{h}_k \cdot \vec{\sigma}$ and define $\mathbb{Z}_2-$topological invariant as $ \nu = \hat{h}_0 \cdot \hat{h}_\pi = \pm 1 $ (where $\hat{h}_k = \frac{\vec{h}_k}{|\vec{h}_k|}$), with $+1$ corresponding to a trivial phase ($|\mu|>t$) and $-1$ to a topological phase ($|\mu|<t$).

\section{Non-Hermitian Kitaev chain}\label{sec:non_herm_kitaev}

We now consider a non-Hermitian extension of the Kitaev chain, Eq.~\eqref{eq:herm_kitaev} and, modify the original Hamiltonian by introducing non-reciprocal hopping parameters and asymmetric superconducting pairing terms, as illustrated in Fig.~\ref{fig:non_herm_schematic}. The Hamiltonian for the non-Hermitian Kitaev chain is given as follows,
\begin{align}\label{eq:non_herm_kitaev}
    H^{\text{NH}} = -\mu \sum_{j=1}^{L} &c_j^{\dagger}c_j - \frac{1}{2} \sum_{j=1}^{L} (t_1 c_j^{\dagger}c_{j+1} + t_2 c_{j+1}^{\dagger}c_j \notag \\ + &\Delta_1 e^{-i\varphi_{1}} c_j c_{j+1} + \Delta_2 e^{i\varphi_{2}}c_{j+1}^{\dagger}c_j^{\dagger} )\,,
\end{align}
where left hopping ($t_{1}$) and right hopping ($t_{2}$) amplitudes are different and also the superconducting pairing terms $\Delta_{1}e^{-i\varphi_{1}}$ and $\Delta_{2}e^{i\varphi_{2}}$ where $\Delta_{1},\,\Delta_{2}\in \mathbb{R}$.

For an infinite chain, we can perform a Fourier transform and obtain the Hamiltonian in the momentum space $H^{\text{NH}}=\frac{1}{2}\sum_k \mathcal{D}_k^{\dagger} \mathcal{H}_{k}^{\text{NH}} \mathcal{D}_{k}$, where,
\begin{align}\label{eq:non_herm_kitaev_momentum}
    \mathcal{H}_k^{\text{NH}} &= \begin{pmatrix}
        -t_1 e^{-ik} - t_2 e^{ik} - 2\mu & 2i\Delta_2 e^{i\varphi_{2}}\sin k \\
        -2i\Delta_1 e^{-i\varphi_{1}}\sin k & t_1 e^{ik} + t_2 e^{-ik} + 2\mu
    \end{pmatrix}\,,
\end{align}
and $\mathcal{D}_{k}^{\dagger} = \begin{pmatrix}
    d_{k}^{\dagger} & d_{-k}
\end{pmatrix}$. 
Diagonalizing the momentum space Hamiltonian Eq.~\eqref{eq:non_herm_kitaev_momentum}, we obtain the two complex energy bands,
\begin{align}\label{eq:non_herm_kspace_spectrum}
    E_{\pm}&(k) = i (t_2 - t_1)\sin k \notag \\ & \pm \sqrt{4\Delta_1 \Delta_2 e^{i(\varphi_{2}-\varphi_{1})} \sin^2 k + \big[(t_1 + t_2) \cos k + 2\mu\big]^2}\,.
\end{align}
For simplicity, we assume that the parameters $\mu, t_1, t_2$ are real and $\varphi_{1} = \varphi_{2} =0$. In this case, we find that the spectrum, Eq.~\eqref{eq:non_herm_kspace_spectrum} is gapless whenever $k=0,\ \pm\pi$ and $\mu = \pm\frac{1}{2}(t_1 + t_2)$. Hence, for an infinite chain, the gap closes at $|\mu| = \tfrac{1}{2}(t_1 + t_2)$, marking the topological phase transition for non-Hermitian Kitaev chain.

\subsection{Finite non-Hermitian Kitaev chain}\label{subsec:finite_non_herm}
\begin{figure}[t!]
    \centering
    \includegraphics[width=1.0\linewidth]{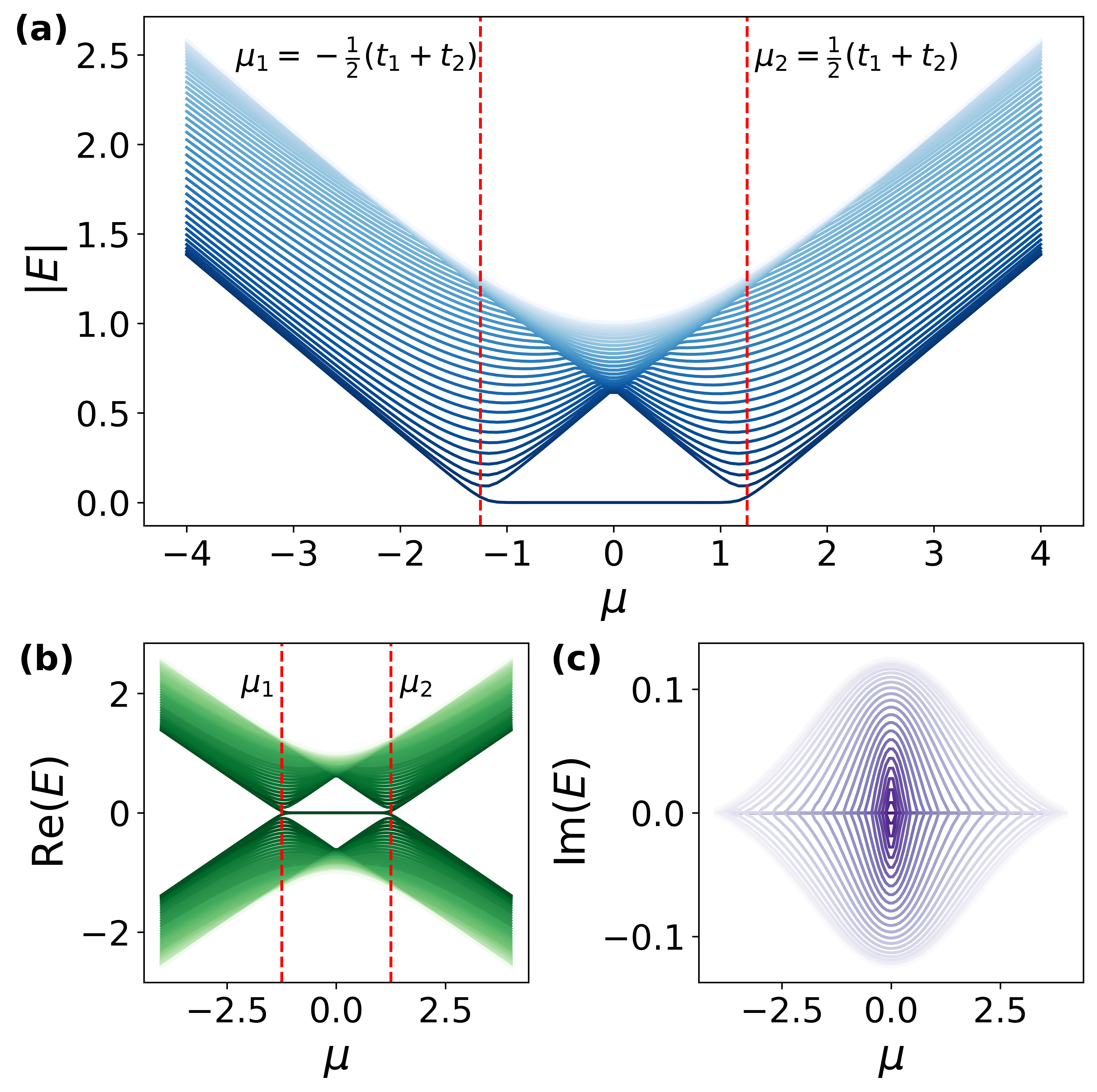}
    \caption{Energy bands of the finite ($L=50$) non-Hermitian Kitaev chain with open boundary conditions (OBC) as function of the chemical potential, $\mu$. Non-Hermiticity is introduced by choosing non-reciprocal hopping parameters i.e., left hopping parameter $t_1 = 1$ and right hopping parameter, $t_{2}=1.5$. The superconducting pairing terms are $\Delta_{1} = 2 = \Delta_{2}$ and $\varphi_{1}=0=\varphi_{2}$ cf., Eq.~\eqref{eq:non_herm_kitaev}. \textbf{(a)} Absolute value of the energy bands as we vary the chemical potential $\mu$. The zero modes (or the topological phase) appear whenever the chemical potential is fine tuned to lie between $\mu_{1}=-\frac{1}{2}(t_{1}+t_{2})$ and $\mu_{2}=\frac{1}{2}(t_{1}+t_{2})$. As a result of breaking Hermiticity, one obtains in general, a complex energy spectrum i.e. non-zero real part \textbf{(b)} and non-zero imaginary part \textbf{(c)} of the energy eigenvalues.}   
    \label{fig:energy_bands_non_herm_kitaev}
\end{figure}

We now analyze the spectrum of a finite chain with open boundary conditions (OBC). The Hamiltonian of a finite chain with $L$ lattice sites with OBC is represented by a $2L \times 2L$ matrix as follows,
\begin{align}\label{eq:finite_non_her_matrix}
    H^{\text{NH}} = \frac{1}{2} &\Psi^{\dagger} \begin{pmatrix}
        A & B & 0 & 0 & 0 & \cdots \\
        C & A & B & 0 & 0 & \cdots \\
        0 & C & A & B & 0 & \cdots \\
        0 & 0 & C & A & B & \cdots \\
        \vdots & \vdots  & \vdots  & \vdots & \vdots & \ddots \\
    \end{pmatrix} \Psi, 
\end{align}
where $\Psi^{\dagger} = \begin{pmatrix}
        c_1^{\dagger} & c_1 & c_2^{\dagger} & c_2 & ...
    \end{pmatrix}$ and $A$, $B$ and $C$ are $2 \times 2$ matrices,
\begin{align}
    A = \begin{pmatrix}
        - \mu & 0 \\
        0 & \mu
    \end{pmatrix} \, \,  B = \begin{pmatrix}
        -\frac{t_1}{2} & \frac{\Delta_2}{2} \\ \notag
        -\frac{\Delta_1}{2} & \frac{t_2}{2}
    \end{pmatrix} \, \,
    C = \begin{pmatrix}
        -\frac{t_2}{2} & -\frac{\Delta_2}{2} \\
        \frac{\Delta_1}{2} & \frac{t_1}{2}
    \end{pmatrix},
\end{align}
where we have again assumed for simplicity $\mu, t_1, t_2 \in \mathbb{R}$ and $\varphi_1 = \varphi_2 = 0$. Diagonalizing the Hamiltonian in Eq.~\eqref{eq:finite_non_her_matrix} reveals a complex band structure, Fig.~\ref{fig:energy_bands_non_herm_kitaev}(b)-(c) and the presence of zero-energy modes in certain parameter regime, Fig.~\ref{fig:energy_bands_non_herm_kitaev}(a).

\begin{figure}[t!]
    \centering
    \includegraphics[width=1.0\linewidth]{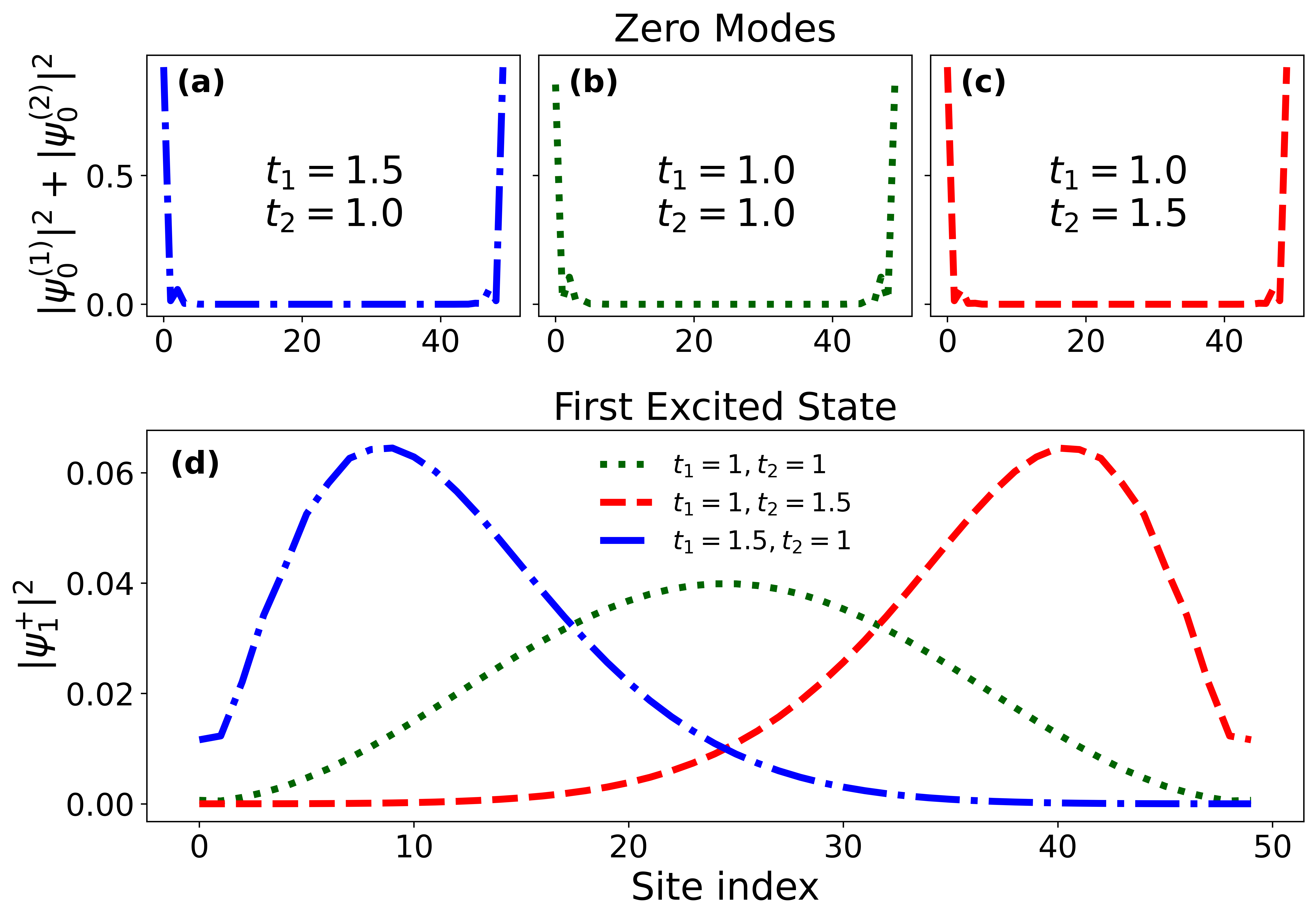}
    \caption{Probability density of the zero modes (doubly degenerate ground states) and first excited state in non-Hermitian Kitaev chain of length $L=50$, with open boundary conditions (OBC) in the topological phase i.e., $|\mu| <\frac{1}{2}(t_{1}+t_{2})$. The chemical potential is $\mu=0.2$ and the superconducting pairing terms are set to $\Delta_{1} = 2 =\Delta_{2}$ with $\varphi_{1}=0=\varphi_{2}$. The zero modes ($\psi_{0}^{(1)}$ and $\psi_{0}^{(2)}$) do not show any skin-effect with non-reciprocal hoppings. However, the first excited state ($\psi_{1}^{+}$) shows skin-effect (right or left localization) depending on the hopping parameters. \textbf{(a)} Sum of probability densities of zero mode wave functions, $\psi_{0}^{(1)}$ and $\psi_{0}^{(2)}$ with hopping parameters $t_{1} = 1.5$ (left hopping), $t_{2}=1$ (right hopping). \textbf{(b)} Sum of probability densities of zero mode wave functions with reciprocal hoppings $t_{1}= 1=t_{2}$. \textbf{(c)} Sum of probability densities of zero mode wave functions with reversed reciprocities in hopping i.e., $t_{1}=1$ and $t_{2}=1.5$. \textbf{(d)} Probability density of the first excited state $\psi_{1}^{+}$ with smallest non-vanishing energy i.e., $|E|\neq 0$ and $\text{Re}(E)>0$, showing left localization (blue) with $t_{1} (=1.5)>t_{2}(=1)$, right localization (red) with $t_{1} (=1)<t_{2}(=1.5)$ in contrast to the case with reciprocal hoppings $t_{1}=1=t_{2}$ (green).}
    \label{fig:wavefunctions}
\end{figure}

The band structure shows that the zero mode solutions only appear for certain values of the chemical potential $\mu$, see Fig.~\ref{fig:energy_bands_non_herm_kitaev}(a), to be precise, the zero modes exist whenever we have $|\mu| < \frac{1}{2}(t_{1}+t_{2})$. This implies that the topological transition point derived using the spectrum of infinite chain, Eq.~\eqref{eq:non_herm_kspace_spectrum} coincides with the transition point obtained from the band structure of finite chain with OBC, cf. Fig.~\ref{fig:energy_bands_non_herm_kitaev}(a)-(b). In order to gain further insight, we look at the wave functions for the finite chain with OBC. With non-reciprocal hoppings ($t_{1}\neq t_{2}$), one would naively expect to see the non-Hermitian skin effect (NHSE)~\cite{PhysRevLett.121.086803}. However, we find that the zero modes do not show NHSE and remain robust to non-reciprocity, Fig.~\ref{fig:wavefunctions}(a)-(c). The NHSE only appears when we consider wavefunctions corresponding to excited states, Fig.~\ref{fig:wavefunctions}(d). In section~\ref{subsec:gbz} and section~\ref{subsec:topological_inv}, we present an explanation for the robustness of the zero modes to NHSE, Fig.~\ref{fig:wavefunctions}(a)-(c), the presence of skin effect in excited states, Fig.~\ref{fig:wavefunctions}(d).

\subsection{Generalized Brillouin zone}\label{subsec:gbz}

The NHSE is an exponential localization of wave functions and is the result of a modified Brillouin zone (BZ) in non-Hermitian systems~\cite{PhysRevLett.121.086803}. The simplest way to obtain the modified Brillouin is to make the substitution $e^{ika}\to \beta$ (note that $a$ is the lattice constant and is set to one), where $|\beta|$ is determined by the non-reciprocities (or the non-Hermiticity) in the hopping parameters~\cite{PhysRevLett.121.086803, PhysRevLett.123.246801}.
Hamiltonian $H^{\text{NH}}$, Eq.~\eqref{eq:non_herm_kitaev_momentum} with the substitution $e^{ik}\to \beta$ becomes,
\begin{equation}\label{eq:non_herm_kitaev_beta}
    H^{\text{NH}}(\beta) = \begin{pmatrix}
        -t_1 \beta^{-1} - t_2 \beta - 2\mu & \Delta_2 e^{i\varphi_2} (\beta - \beta^{-1}) \\ 
        \Delta_1 e^{-i\varphi_1} (\beta^{-1} - \beta) & t_1 \beta + t_2 \beta^{-1} + 2\mu
    \end{pmatrix}.
\end{equation}
 
With the modified Hamiltonian Eq.~\eqref{eq:non_herm_kitaev_beta}, we can solve the eigenvalue equation, $\text{det}(H^{\text{NH}}(\beta) - E)=0$ to obtain the energy eigenvalues as a function of $\beta$. We are interested in understanding the behavior of the zero modes of the modified Hamiltonian, Eq.~\eqref{eq:non_herm_kitaev_beta}. For $E=0$, we have, $\text{det}H^{\text{NH}}(\beta) = 0$ and hence we have the following,
\begin{align}\label{eq:beta_non_herm_eqn}
    A_{2}\big[ \beta^{2} + \frac{1}{\beta^{2}} \big] +A_{1}\big[ \beta + \frac{1}{\beta} \big] + A_{0} = 0
\end{align}
where
\begin{align}
    A_{2} &= \Delta_{1}\Delta_2 e^{i(\varphi_2 - \varphi_1)} - t_{1}t_{2}\notag \\ 
    A_{1} &= - 2\mu(t_{1}+t_{2})\notag \\
    A_{0} &= - (t_{1}^{2}+t_{2}^{2}+4\mu^{2}+2\Delta_{1}\Delta_{2} e^{i(\varphi_2 - \varphi_1)}).
\end{align}
From Eq.~\eqref{eq:beta_non_herm_eqn} we find that if we have $\beta_{0}$ as a solution, then $\beta_{0}^{-1}$ is also a solution. Hence, in general, we have four solutions to Eq.~\eqref{eq:beta_non_herm_eqn}: $\beta_1,\,\beta_1^{-1},\,\beta_2,\,\beta_2^{-1}$. Additionally, one can show that the matrix, Eq.~\eqref{eq:non_herm_kitaev_beta} can be null only at the boundary of the topological and trivial phase (this can be seen directly from Eq. \eqref{eq:non_herm_kitaev_beta}, which becomes null only at $\beta=\pm 1$ and $\mu = \mp \frac{t_1 + t_2}{2}$), hence the rank of the matrix in Eq. \eqref{eq:non_herm_kitaev_beta} is 1 at all points within the topological phase. Hence, corresponding to each $\beta$, there is only one zero mode eigenvector in the topological phase. Thus, the four roots of Eq.~\eqref{eq:beta_non_herm_eqn} will have four zero mode eigenvectors in the topological phase. Specifically, at $\mu = 0$, we have $\beta_2 = -\beta_1$. On physical grounds, we expect that the zero mode eigenvectors corresponding to $\beta_1 (\beta_1^{-1})$ will be the same as $-\beta_1(-\beta_1^{-1})$. Consequently, for $E=0$, the two solutions have inverse localization factors $\beta_1$ and $\beta_1^{-1}$, which means that the solution $\beta_1$ ($\beta_1^{-1}$) localizes to the right (left)~\cite{Li_2023, Yan_2023,https://doi.org/10.1002/pssb.202200448}. The total probability density of the zero-energy states is given by $|\psi_0^{(1)}|^2 + |\psi_0^{(2)}|^2$, where $\psi_0^{(1)}$ and $\psi_0^{(2)}$ are the normalized zero modes corresponding to factors $\beta_1$ and $\beta_1^{-1}$ respectively. This reciprocal pairing eliminates the skin effect for zero-energy states, as seen in Fig. \ref{fig:wavefunctions}(a)-(c). The reciprocal solution pair for Eq.~\eqref{eq:beta_non_herm_eqn} is a result of particle-hole symmetry in the non-Hermitian Kitaev chain~\cite{Li_2023}. 
However, for any excited state or bulk mode, the solutions to $\text{det}(H^{\text{NH}}(\beta) - E)=0$ are not in a reciprocal pair. In fact, if $\beta_{1}$ is a solution corresponding to energy $E$, then $\beta_{1}^{-1}$ is a solution corresponding to energy $-E$. Hence, eigenstates with energies $\pm E$ exhibit equal and opposite non-Hermitian skin effect, as shown in Fig.~\ref{fig:complex_parameters_skin_effect}, and also see appendix~\ref{AppendixA} for more details.

The solutions for $\beta$ in Eq.~(\ref{eq:beta_non_herm_eqn}) encode the localization length of the Majorana zero modes. This can be understood in a simple setting by considering $\mu=0$ and setting $\varphi_1=\varphi_2$. As a result Eq.~(\ref{eq:beta_non_herm_eqn}) reduces to a quadratic equation for $\beta^2$ with reciprocal roots,
\begin{equation}\label{eq:beta_pm}
    \beta^2_{\pm}=\frac{t_1^2 + t_2^2 + 2\Delta_1 \Delta_2 \pm \sqrt{(t_1^2 - t_2^2)^2 + 4(t_1 + t_2)^2 \Delta_1 \Delta_2}}{2(\Delta_1 \Delta_2 - t_1 t_2)}.
\end{equation}
In non-Hermitian lattices with skin effect, the wavefunctions go as $\psi_j \sim \beta^j$~\cite{PhysRevLett.121.086803}, and the localization length is given by $a/\big{|}\log |\beta|\big{|}$, where $a$ is the lattice constant which is set to 1 for simplicity. Thus, the localization length for the two Majorana zero modes are $1/\big{|}\log |\beta_+|\big{|}$ and $1/\big{|}\log |\beta_-|\big{|}$, which are equal since we have $\beta_{+} \beta_{-} = 1$. Thus, the zero modes are always exponentially localized at the two edges of the chain. In the Hermitian case where $t_1=t_2=t$ and $\Delta_1=\Delta_2=\Delta$ in Eq.~(\ref{eq:beta_pm}) we find $1/\big{|}\log |(\Delta+t)/(\Delta-t)|\big{|}$ as the localization length for the Majorana modes as expected from previous analysis~\cite{Alicea_2012}.

Next, we discuss the role of $\Delta_1$ and $\Delta_2$ in the physics of the non-Hermitian Kitaev chain, especially in the robustness of the Majorana modes. One might expect that setting $\Delta_{1}$ and $\Delta_{2}$ arbitrarily small reduces the non-Hermitian Kitaev chain to the Hatano-Nelson model, where all states exhibit the skin effect. However note that the role of even arbitrarily small superconducting pairing terms in the Hamiltonian Eq.~\eqref{eq:herm_kitaev} and Eq.~\eqref{eq:non_herm_kitaev} is to open a gap in the spectrum. The closing and reopening of this bulk gap drives the system from the topological phase (with Majorana zero modes at the edges) to the trivial phase (without zero modes), and vice-versa. As discussed previously, Majorana modes are exponentially localized at the edges of the chain as long as $\Delta_{1,2} \neq 0$. In contrast, the Hatano-Nelson model does not have a topological phase transition.

The Hamiltonian of the non-Hermtian Kitaev chain with $\Delta_{1,2}\neq 0$ takes a Bogoliubov-de-Gennes (BdG) form, Eq.~\eqref{eq:non_herm_kitaev_momentum}, which is another manifestation of the particle-hole symmetry in the system. The BdG form holds for arbitrary complex parameters, which implies an unbroken paricle-hole symmetry throughout the entire complex parameter regime, as we shall demonstrate explicitly in the following subsection~\ref{subsec:sym_non_herm_kitaev}. Hence, we see that $\Delta_{1,2}\neq 0$ makes the system very different from the Hatano-Nelson model, and leads to significantly different physics. As a result of particle-hole symmetry, we get reciprocal roots for $\beta$, leading to an equal and opposite skin effect for particles at energy $+E$ and holes at energy $-E$. As Majorana zero modes are an equal superposition of particles and holes at zero energy, the skin effect is neutralized for these modes, and we get symmetrically localized Majorana zero modes at the edges, cf. Fig.~\ref{fig:wavefunctions}(a)-(c).

\subsection{Symmetry analysis of non-Hermitian Kitaev chain}\label{subsec:sym_non_herm_kitaev}

Our discussion so far has been focused on a simplified version of the non-Hermitian Kitaev chain, i.e., keeping the parameters $t_{1}$, $t_{2}$, $\mu$ real and setting $\varphi_{1}=\varphi_{2}=0$. We now study the non-Hermitian Kitaev chain by relaxing this constraint. With complex parameters, the analysis done so far becomes cumbersome. However, understanding the symmetries of the system makes the analysis easier. Here, following Ref.~\cite{Kawabata_2022}, we study the symmetries of the non-Hermitian Kitaev chain.

\begin{table}[b!]
\centering
\begin{tabular}{|c | c | c | c | c | c | c|} 
 \hline
 Condition & TRS & PHS & TRS${}^{\dagger}$ & PHS${}^{\dagger}$ & CS & CS$^{\dagger}$ \\ [0.5ex] 
 \hline\hline
 $t_{1}=t_{2}$\,,  $\varphi_{1} = \varphi_{2}$  & \cmark & \cmark  & \cmark & \cmark  & \cmark &  \cmark\\ 
 \hline
 $t_{1}=t_{2}$\,, $\varphi_{1} \neq \varphi_{2}$ & \xmark & \cmark & \cmark & \xmark & \xmark & \cmark\\
 \hline
 $t_{1}\neq t_{2}$\,, $\varphi_{1} = \varphi_{2}$ & \cmark & \cmark & \xmark & \xmark & \cmark & \xmark \\
 \hline
 $t_{1}\neq t_{2} $\,, $\varphi_{1} \neq \varphi_{2}$ & \xmark & \cmark & \xmark & \xmark & \xmark &  \xmark \\
 \hline
 $t_{1}=t_{2}\in\mathbb{C} \backslash \mathbb{R} \, $, $\varphi_{1} = \varphi_{2}$ & \xmark & \cmark & \cmark & \xmark & \xmark & \cmark \\ 
 \hline
 $t_{1}^{*}= t_{2}\in\mathbb{C} \backslash \mathbb{R} \, $, $\varphi_{1} = \varphi_{2}$ & \xmark & \cmark & \xmark & \cmark &  \xmark & \xmark \\ 
 \hline
 $t_{1}\neq t_{2}\in\mathbb{C} \backslash \mathbb{R} \,$, $\varphi_{1} = \varphi_{2}$& \xmark & \cmark & \xmark & \xmark &  \xmark & \xmark \\ 
 \hline
 $t_{1}=t_{2}\in \mathbb{C} \backslash \mathbb{R} \,$, $\varphi_{1} \neq \varphi_{2}$ & \xmark & \cmark & \cmark & \xmark & \xmark  & \cmark \\ 
 \hline
 $t_{1}^{*}=t_{2}\in\mathbb{C} \backslash \mathbb{R} \, $, $\varphi_{1} \neq \varphi_{2}$ & \xmark & \cmark & \xmark & \cmark & \xmark & \xmark \\
 \hline
 $t_{1}\neq t_{2}\in\mathbb{C} \backslash \mathbb{R} \,$, $\varphi_{1} \neq \varphi_{2}$ & \xmark & \cmark & \xmark & \xmark & \xmark &\xmark  \\  
 \hline
\end{tabular}
\caption{Symmetries of the non-Hermitian Kitaev chain for general set of parameters. Hopping amplitudes $t_{1}$ and $t_{2}$ are chosen to be real ($t_{1},\,t_{2}\in\mathbb{R}$) unless specified otherwise. We also restrict ourselves to $\mu \in \mathbb{R}$ in this table. It is important to note that the superconducting pairing amplitudes $\Delta_1 , \Delta_2$ do not affect the symmetries and therefore are unconstrained. Further, we also see that the particle hole symmetry is always present in the system and is the reason for the robustness of the Majorana zero-modes to the non-Hermitian skin effect.}
\label{tab:sym_tab}
\end{table}

In contrast to Hermitian systems, where the condition $H^\dagger = H$ (for Hamiltonian) ties together transpose and complex-conjugation operations, non-Hermitian Hamiltonians require that these operations be treated as independent when defining discrete symmetries. For example, when studying time-reversal symmetry (TRS) in a non-Hermitian system, one should address $H^{*}$ and $H^{T}$ separately, corresponding to TRS and its adjoint TRS$^{\dagger}$. Similar statements hold for chiral-symmetry (CS) and particle-hole symmetry (PHS)~\cite{Kawabata_2022}. Considering arbitrary complex parameters, cf. Eq.~\eqref{eq:non_herm_kitaev} and by using Specht's theorem~\cite{Horn_Johnson_1985}, we summarize the symmetries of the non-Hermitian Kitaev chain in Table~\ref{tab:sym_tab}.

\begin{figure}[t!]
    \centering
    \includegraphics[width=\linewidth]{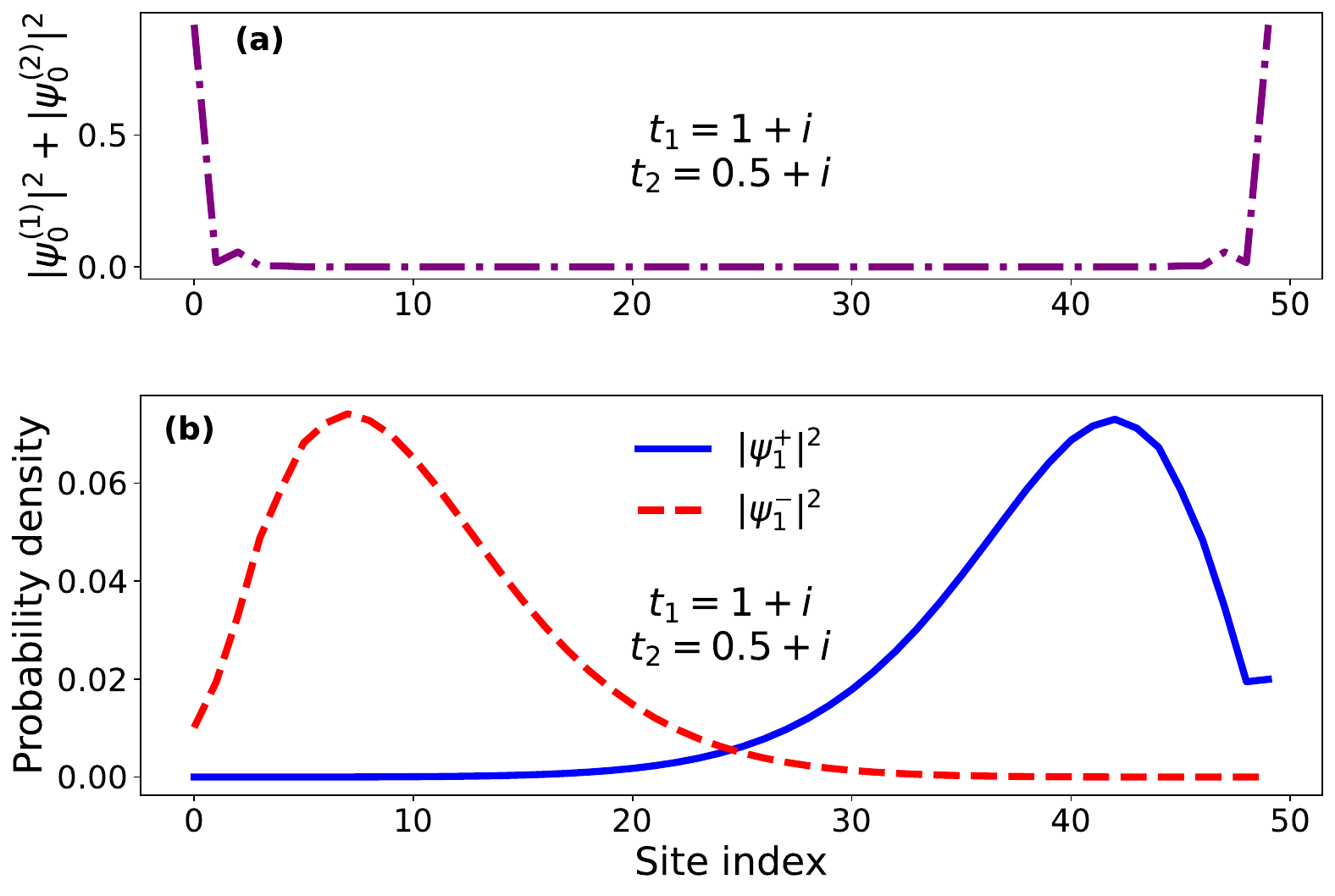}
    \caption{Probability density of zero mode wavefunctions and eigenstates with smallest non-vanishing energy ($\pm |E| \neq 0$), of the non-Hermitian Kitaev chain with $L=50$, $\mu=0.2$, $t_1 =1+i$, $t_2=0.5+i$, $\Delta_1=2+i$ and $\Delta_2=1+i$. The zero modes are immune to the NHSE whereas the nonzero energy wavefunctions show an equal and opposite skin effect. \textbf{(a)} Sum of probability densities of the zero modes $\psi_0^{(1)}$ and $\psi_0^{(2)}$ is symmetric and localized at the edges. \textbf{(b)} $\psi_1^+$ and $\psi_1^-$ are the energy eigenstates corresponding to complex energies $\pm E$, $\text{Re}(E)>0$ such that $|E|$ is the smallest non-zero absolute value of energy in the spectrum i.e., $|E|\neq 0$. The solid blue line and the dashed red lines denote the probability densities of $\psi_1^+$ and $\psi_1^-$ respectively and exhibit opposite skin effect.}
    \label{fig:complex_parameters_skin_effect}
\end{figure}

It is important to note that the particle-hole symmetry (PHS) always persists, with arbitrary complex parameters. The Table~\ref{tab:sym_tab} presented here considers real $\mu$. For complex $\mu$, PHS still holds, and the only other possible symmetries are TRS$^{\dagger}$ and CS$^{\dagger}$, see Appendix~\ref{appendix}. As a result of PHS, the Majorana zero modes exist and are robust across the entire parameter regime, Fig.~\ref{fig:wavefunctions}(a)-(c) and Fig.~\ref{fig:complex_parameters_skin_effect}(a). However, the bulk modes exhibit NHSE with wavefunctions of opposite energies localizing at opposite edges, Fig.~\ref{fig:complex_parameters_skin_effect}(b). To see this, note that PHS on the energy spectrum implies that $E(k)$ and $-E(-k)$ come in pairs. Once we promote $e^{ik}$ to $\beta$, this implies that $E(\beta)$ and $-E(\beta^{-1})$ come in pairs, i.e., the states with energy $+E$ and $-E$ localize at opposite ends. We emphasize that Fig.~\ref{fig:complex_parameters_skin_effect} (and Fig.~\ref{fig:supp_fig1} in Appendix~\ref{AppendixA}) demonstrates that the symmetric localization of the Majorana zero modes and the equal and opposite skin effect for eigenstates with energy $\pm E$ for arbitrarily complex parameters in comparison to Fig.~\ref{fig:wavefunctions}, where we had $\Delta_1 = \Delta_2$, and all real parameters.

\subsection{Topological phase diagram for non-Hermitian Kitaev chain}\label{subsec:topological_inv}

We have seen that the topological phase transition point coincides for a finite open chain and an infinite or periodic chain in the non-Hermitian Kitaev chain. Additionally, we found that the zero-mode solution exists for a range of values of the chemical potential $\mu$, see Fig.~\ref{fig:energy_bands_non_herm_kitaev}. Hence, it is important to identify the topological and trivial phase in the non-Hermitian Kitaev chain, Eq.~\eqref{eq:non_herm_kitaev} as a function of the parameters. Therefore, we would like to have a topological invariant that can detect the phase, similar to the Hermitian version, cf. section~\ref{sec:herm_kitaev}.

Similar to the Hermitian Kitaev chain~\cite{Alicea_2012}, we begin by writing our Hamiltonian, Eq.~\eqref{eq:non_herm_kitaev_momentum} in the Pauli basis i.e., $H(k) = h_\mathbbm{1}(k)\mathbbm{1} + h_x(k)\sigma_x + h_y(k)\sigma_y + h_z(k)\sigma_z$, where $\mathbbm{1}$ is the identity matrix and $\sigma_{x},\,\sigma_{y},\,\sigma_{z}$ are the Pauli matrices. The coefficients $h_{\mathbbm{1}}(k),\,h_{x}(k),\,h_{y}(k)$ and $h_{z}(k)$ are given as follows,
\begin{align}
    h_\mathbbm{1}(k) &= i(t_1 - t_2) \sin k \,,\notag \\
    h_x(k) &= i(\Delta_2 e^{i\varphi_{2}} - \Delta_1 e^{-i\varphi_{1}}) \sin k \,, \notag \\
    h_y(k) &= -(\Delta_2 e^{i\varphi_{2}} + \Delta_1 e^{-i\varphi_{1}}) \sin k \notag \,, \\ 
    h_z(k) &= -(t_1 \cos k + t_2 \cos k + 2\mu).
\end{align}
We begin with the simple case of real chemical potential ($\mu \in \mathbb{R}$), real hopping parameters ($t_{1},\,t_{2} \in \mathbb{R}$) and superconducting phases $\varphi_1 = \varphi_2 = 0$. We note that the gap in the bulk spectrum, Eq.~\eqref{eq:non_herm_kspace_spectrum} can only close at the points $k=0,\,\pm \pi$, where we have $h_{z}(k=0) = -(t_{1}+t_{2}+2\mu)$ and $h_{z}(k=\pm\pi) = (t_{1}+t_{2}-2\mu)$. We can compactly define $\frac{[h_z(0)]}{|t_{1}+t_{2}+2\mu|}=-\text{sgn}\{t_{1}+t_{2}+2\mu\}$ and $\frac{[h_z(\pm \pi)]}{|t_{1}+t_{2}-2\mu|}=\text{sgn}\{t_{1}+t_{2}-2\mu\}$. This enables us to define the topological invariant in this simple case as follows,
\begin{align}\label{eq:z2invariant}
    \nu = \frac{[h_z(0)]}{|t_{1}+t_{2}+2\mu|}\cdot \frac{[h_z(\pm \pi)]}{|t_{1}+t_{2}-2\mu|} \,.
\end{align}
For $t_{1} + t_{2} \geq 2|\mu|$, we have $\nu = -1$, which is identified with the topological phase with two zero modes. In the case of $t_{1} + t_{2} < 2|\mu|$, we have $\nu= +1$, which corresponds to the trivial phase, cf. Fig.~\ref{fig:energy_bands_non_herm_kitaev}(a)-(c). In the most general parameter regime, $t_1 \neq t_2, t_1 \neq t_2^*, \varphi_1 \neq \varphi_2$, PHS is the only symmetry of the non-Hermitian Kitaev chain (refer Table~\ref{tab:sym_tab}). Using the symmetry-based topological classification of Ref.~\cite{PhysRevX.9.041015}, we observe that the non-Hermitian Kitaev chain is in class D, and hence has a $\mathbb{Z}_2$ topological invariant, see Appendix \ref{appendix} for more details. It is also worth noting that in experimental setups, the physical observable to detect Majorana zero modes appearing at the ends of a superconducting wire is fermion parity \cite{Aghaee2025}, which also corresponds to a $\mathbb{Z}_2$ topological invariant. 

\begin{figure}
    \centering
    \includegraphics[width=1.0\linewidth]{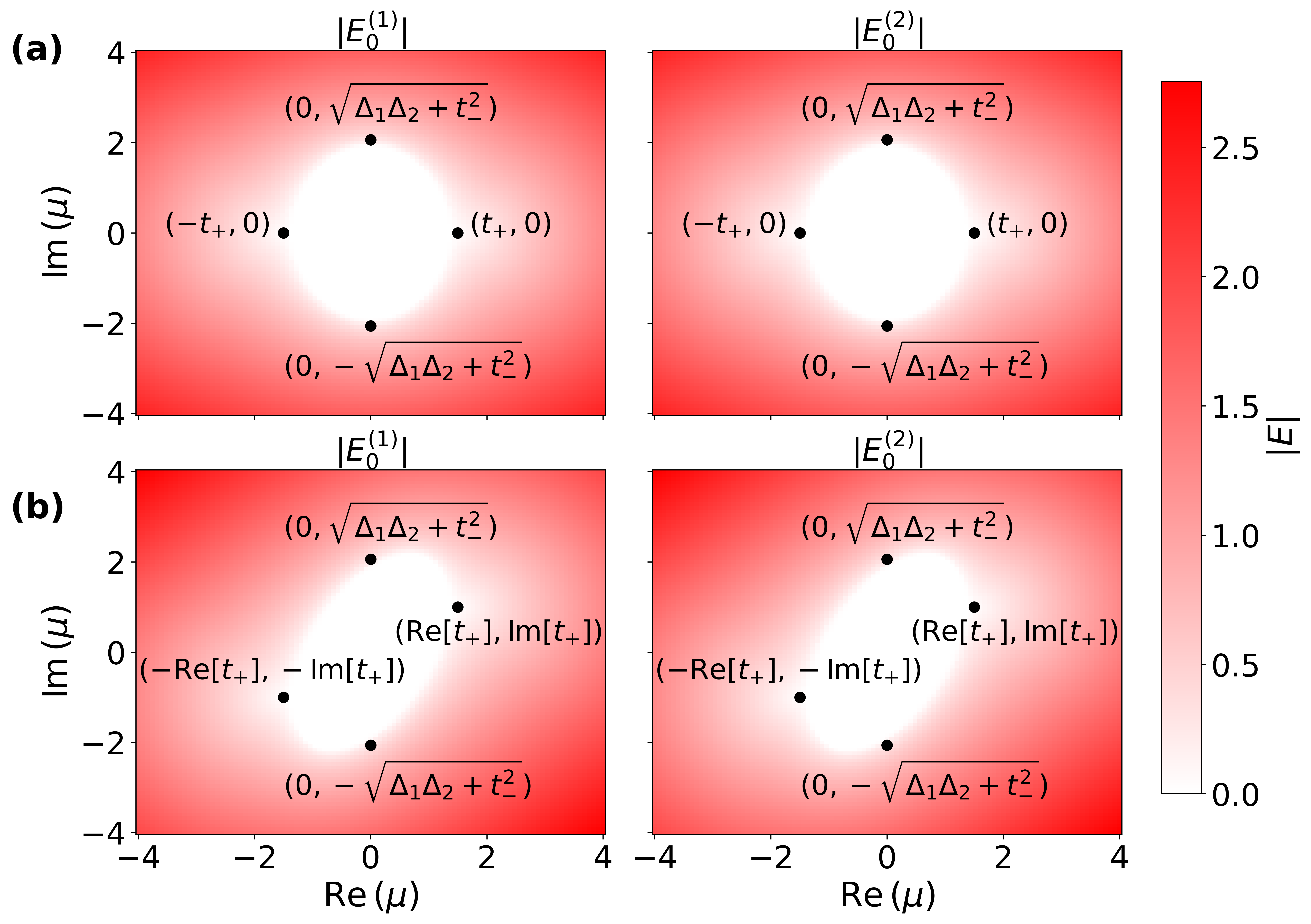}
    \caption{Topological phase diagram for non-Hermitian Kitaev chain in the complex chemical potential ($\mu$) plane. The quantities $E_{0}^{(1)}$ and $E_{0}^{(2)}$ are the two lowest eigen-energies (absolute values) of the non-Hermitian Kitaev chain, Eq.~\eqref{eq:non_herm_kitaev}. Vanishing of the two (eigen-energies, $E_{0}^{(1)}$ and $E_{0}^{(2)}$) corresponds to the Majorana-edge modes and hence marks the topological phase. \textbf{(a)} Phase diagram with non-reciprocal real hopping parameters. For obtaining the phase diagram we have used the following specific parameters, $t_{1}=1,\,t_{2}=2, \,\Delta_{1}=2,\,\Delta_{2} = 2, \,\text{and} \,\varphi_{1}=0=\varphi_{2}$. \textbf{(b)} Phase diagram with non-reciprocal complex hopping parameters with equal imaginary parts i.e., $\text{Im}(t_{1})=\text{Im}(t_{2})$. Specific choice of parameters for obtaining the phase diagram are as follows, $t_{1}=1 + i,\,t_{2}=2 + i, \,\Delta_{1}=2,\,\Delta_{2} = 2, \,\text{and} \,\varphi_{1}=0=\varphi_{2}$.}
    \label{fig:phase_diagram}
\end{figure}

Next, we would like to understand the phase diagram with complex chemical potential ($\mu \in \mathbb{C}$) while keeping $t_{1},\,t_{2}\in \mathbb{R}$ and $\varphi_{1}=\varphi_{2}=0$. This leads to infinite number of gap closing points in the complex $\mu-$plane, Fig.~\ref{fig:phase_diagram} in contrast to only two solutions for real $\mu$. The transition points along the Im($\mu$) axis can be obtained by noting that the gap closing points are at $k=\pm \pi/2$. This gives Im$(\mu)= \pm\sqrt{\Delta_{1}\Delta_{2}+t_{-}^{2}}$, where $t_{-}=\frac{1}{2}(t_{1}-t_{2})$ and as shown in the Fig.~\ref{fig:phase_diagram}(a). Similarly, the transition points along the Re$(\mu)$ axis are obtained by noting the gap closing points are at $k=0,\,\pm \pi$ and gives the transition points at Im$(\mu)= \pm t_{+}$, where $t_{+}=\frac{1}{2}(t_{1}+t_{2})$, Fig.~\ref{fig:phase_diagram}(a). Additionally, with complex hopping amplitudes, it turns out that we can analytically determine four points on the phase boundary for a special case, where $\text{Im}(t_{1}) = \text{Im}(t_{2})$. Gap-closing of the spectrum occurs at $k=0,\,\pm \pi$ and $k=\pm \pi/2$ corresponding to the points $(0,\pm \sqrt{\Delta_{1}\Delta_{2}+t_{-}^{2}})$ and $(\pm\text{Re}[t_{+}],\pm \text{Im}[t_{+}])$, cf. Fig.~\ref{fig:phase_diagram}(b). 

\section{Conclusion}

The analysis presented here focuses on the non-Hermitian Kitaev chain where non-Hermiticity is introduced via non-reciprocal hopping amplitudes ($t_{1}\neq t_{2}$) and asymmetric superconducting pairing terms ($\Delta_{1}e^{-i\varphi_{1}}\neq \Delta_{2}e^{i\varphi_{2}}$), cf. Eq.~\eqref{eq:non_herm_kitaev}. Considering arbitrary complex parameters, we show the existence of different discrete symmetries corresponding to each parameter regime, see Table~\ref{tab:sym_tab}.

We demonstrated that the transition point between the topological and trivial phase with periodic boundary condition (equivalently infinite chain) coincides with the open boundary condition for a finite chain. The underlying reason is the existence of particle-hole symmetry across entire parameter regime, see Table~\ref{tab:sym_tab}. As a result, the solutions to the localization parameter $\beta$, Eq.~\eqref{eq:beta_non_herm_eqn} for the zero modes (doubly degenerate) of the Hamiltonian always exist in reciprocal pairs~\cite{Li_2023,Yan_2023,https://doi.org/10.1002/pssb.202200448} and therefore zero-modes are robust to non-Hermitian skin effect (NHSE), see Fig.~\ref{fig:wavefunctions}(a)-(c) and Fig.~\ref{fig:complex_parameters_skin_effect}(a). We also study the bulk (non-zero energy) modes and find that the particle and hole wavefunctions show equal and opposite skin effect, as implied by PHS, see Fig.~\ref{fig:complex_parameters_skin_effect}(b).

Further, we propose a topological invariant, Eq.~\eqref{eq:z2invariant} that determines the phase (trivial or topological) of a non-Hermitian Kitaev chain with real hopping amplitudes and chemical potential ($t_{1}\,,t_{2}\,,\mu \in \mathbb{R}$). Finally, by relaxing constraints on the parameters, we present phase diagrams for the non-Hermitian Kitaev chain in the complex chemical potential ($\mu$) with real and imaginary hoppings, Fig.~\ref{fig:phase_diagram}.

The interplay of topology and non-Hermiticity is very interesting and gives rise to many new theoretical ideas such as the non-Hermitian skin effect and non-Bloch bulk-boundary correspondence. There are also numerous experimental validations of these ideas in mechanical metamaterials~\cite{doi:10.1073/pnas.2010580117,PhysRevLett.131.207201, s97b-qcjc}, topoelectric circuits~\cite{zou_observation_2021,7dvt-k6gk}, and photonic systems~\cite{doi:10.1126/science.aaz8727,Xiao2020,PhysRevResearch.2.013280,zhang_acoustic_2021,gao_non-hermitian_2021,zhou_observation_2023,wan_observation_2023}. The key idea in all such experimental platforms is not to deal with real quantum Hamiltonians, but only to mimic the physics of a quantum Hamiltonian. This is achieved by tuning various electrical components such as resistors, inductors and capacitors in topoelectric circuits~\cite{zou_observation_2021}, by artificially engineering non-reciprocal hopping amplitudes in active mechanical circuits~\cite{doi:10.1073/pnas.2010580117}, or by modulating gain and loss in unidirectional coupling link rings in optical systems~\cite{PhysRevResearch.2.013280}. Additionally, non-Hermitian physics has also found application in designing new highly sensitive sensors, as demonstrated in Ref.~\cite{doi:10.34133/research.1091}. The versatility of these platforms allows for the engineering of a wide range of non-reciprocal couplings and effective interactions~\cite{https://doi.org/10.1002/lpor.202400099}. Realizing a non-Hermitian Kitaev chain may be challenging because of the superconducting pairing terms. However, with all the advances in the experimental field, it is not unreasonable to be optimistic that such interactions may be realized, or at least the underlying physics may be mimicked in an experimental platform.

Overall, in this work we have provided a comprehensive characterization of the non-Hermitian Kitaev chain over the full complex parameter space, establishing a consistent correspondence between bulk topology, zero-energy Majorana modes, and phase boundaries. For specific parameter regimes, we obtained analytical expressions for points on the topological phase boundary, as illustrated in Fig.~\ref{fig:phase_diagram}, offering concrete insight into the structure of the phase diagram. It remains to be seen whether the phase diagram can be determined analytically in full generality, including a characterization of the shape of the phase boundary; nevertheless, our results lay a systematic and broadly applicable foundation for exploring non-Hermitian topological phase transitions in this and related models.

\section*{A\MakeLowercase{cknowledgements}}

We thank Jukka I. Vayrynen and Qi Zhou for insightful comments and discussions. A.R. would also like to thank Laimei Nie and Eric Schultz for useful discussions. We also thank Sayak Ray for comments on the draft of the manuscript.

\begin{appendix}
\begin{widetext}

\renewcommand{\thesection}{\Alph{section}}
\renewcommand{\theequation}{\thesection\arabic{equation}}
\renewcommand{\thefigure}{\thesection\arabic{figure}}
\renewcommand{\thetable}{\thesection\arabic{table}}

\makeatletter
\@addtoreset{equation}{section}
\@addtoreset{figure}{section}
\@addtoreset{table}{section}
\makeatother

\section{Probability densities for ground and excited states}\label{AppendixA}

In this appendix, we elaborate on obtaining the probability densities of the zero energy modes ($E=0$) and the first excited states (with energy $\pm E$). By first excited state we mean eigenstates with lowest magnitude of energy with $|E|\neq 0$. Consider the Hamiltonian discussed in the main text, Eq.~(\ref{eq:finite_non_her_matrix}) as follows,
\begin{align}\label{eq:finite_non_her_matrix}
    H^{\text{NH}} = \frac{1}{2} &\Psi^{\dagger} \begin{pmatrix}
        A & B & 0 & 0 & 0 & \cdots \\
        C & A & B & 0 & 0 & \cdots \\
        0 & C & A & B & 0 & \cdots \\
        0 & 0 & C & A & B & \cdots \\
        \vdots & \vdots  & \vdots  & \vdots & \vdots & \ddots \\
    \end{pmatrix} \Psi,
\end{align}
where $\Psi^{\dagger} = \begin{pmatrix}
        c_1^{\dagger} & c_1 & c_2^{\dagger} & c_2 & \cdots
    \end{pmatrix}$ and $A$, $B$ and $C$ are $2 \times 2$ matrices,
\begin{align}
    A = \begin{pmatrix}
        - \mu & 0 \\
        0 & \mu
    \end{pmatrix} \qquad  B = \begin{pmatrix}
        -\frac{t_1}{2} & \frac{\Delta_2}{2} \\ \notag
        -\frac{\Delta_1}{2} & \frac{t_2}{2}
    \end{pmatrix} \qquad
    C = \begin{pmatrix}
        -\frac{t_2}{2} & -\frac{\Delta_2}{2} \\
        \frac{\Delta_1}{2} & \frac{t_1}{2}
    \end{pmatrix},
\end{align}
Let us start with an eigenvector of the Hamiltonian corresponding to energy $E$. Denote the $(2j-1)^{\text{th}}$ component of the eigenvector as $u_j$ and $2j^{\text{th}}$ component as $v_j$ where $j \in {1,2,\cdots, L}$. We identify the odd component of the eigenstate $\psi_{1,j} = u_j c_j$ to be the corresponding component for the particle wavefunction and similarly $\psi_{2,j} = v_j c_j^{\dagger}$ is the component of the hole wavefunction. Thus, the probability density of the particle with energy $E$ at $j^{\text{th}}$ site is given by $|u_j|^2$ and for the hole, density at $j^{\text{th}}$ site is $|v_j|^2$.

For the eigenstates of the Hamiltonian corresponding to the energy eigenstates $\pm E$, let us denote the particle and hole probability density at $j^{\text{th}}$ site with $|u^{\pm}(j)|^2$ and $|v^{\pm}(j)|^2$. The particle-hole symmetry (PHS) of the non-Hermitian Kitaev chain implies that if a particle with energy $\pm E$ localizes at one end of the chain, then a hole of energy $\mp E$ localizes at the other end (see Sec. \ref{subsec:sym_non_herm_kitaev}). Now, the probability densities of these two wavefunctions $\psi_{\pm}$ at site $j$ are given by
\begin{equation}
    |\psi_{\pm}(j)|^2 = |u^{\pm}_j|^2 + |v^{\pm}_j|^2.
\end{equation}
As a result of PHS, probability density of particles $|u^+|^2$ with energy $E$, has an equal and opposite localization as compared to the probability density of holes $|v^-|^2$. Equivalently, $|u^-|^2$ has an equal and opposite localization as compared to $|v^+|^2$. This immediately leads to the conclusion that the probability density $|\psi^+|^2$ of the wavefunction corresponding to energy $+E$ has an equal and opposite localization as compared to the probability density $|\psi^-|^2$ of the wavefunction corresponding to energy $-E$ cf. Fig.~\ref{fig:supp_fig1}(b).

In the case of Majorana modes, we present the combined probability density i.e., summing the probability density for both the zero modes as follows,
\begin{equation}
    |\psi_0^{(1)}|^2 + |\psi_0^{(2)}|^2 = |u_0^{(1)}|^2 + |v_0^{(1)}|^2 + |u_0^{(2)}|^2 + |v_0^{(2)}|^2,
\end{equation}
where $u_0^{(1,2)}$ and $v_0^{(1,2)}$ are the particle and hole components corresponding to the two Majorana zero modes $\psi_0^{(1,2)}$. It is now evident that the symmetric profile of the combined probability density of the Majorana zero modes is also a consequence of PHS cf. Fig.~\ref{fig:supp_fig1}(a).

\begin{figure}[t]
    \centering
    \includegraphics[width=0.5\linewidth]{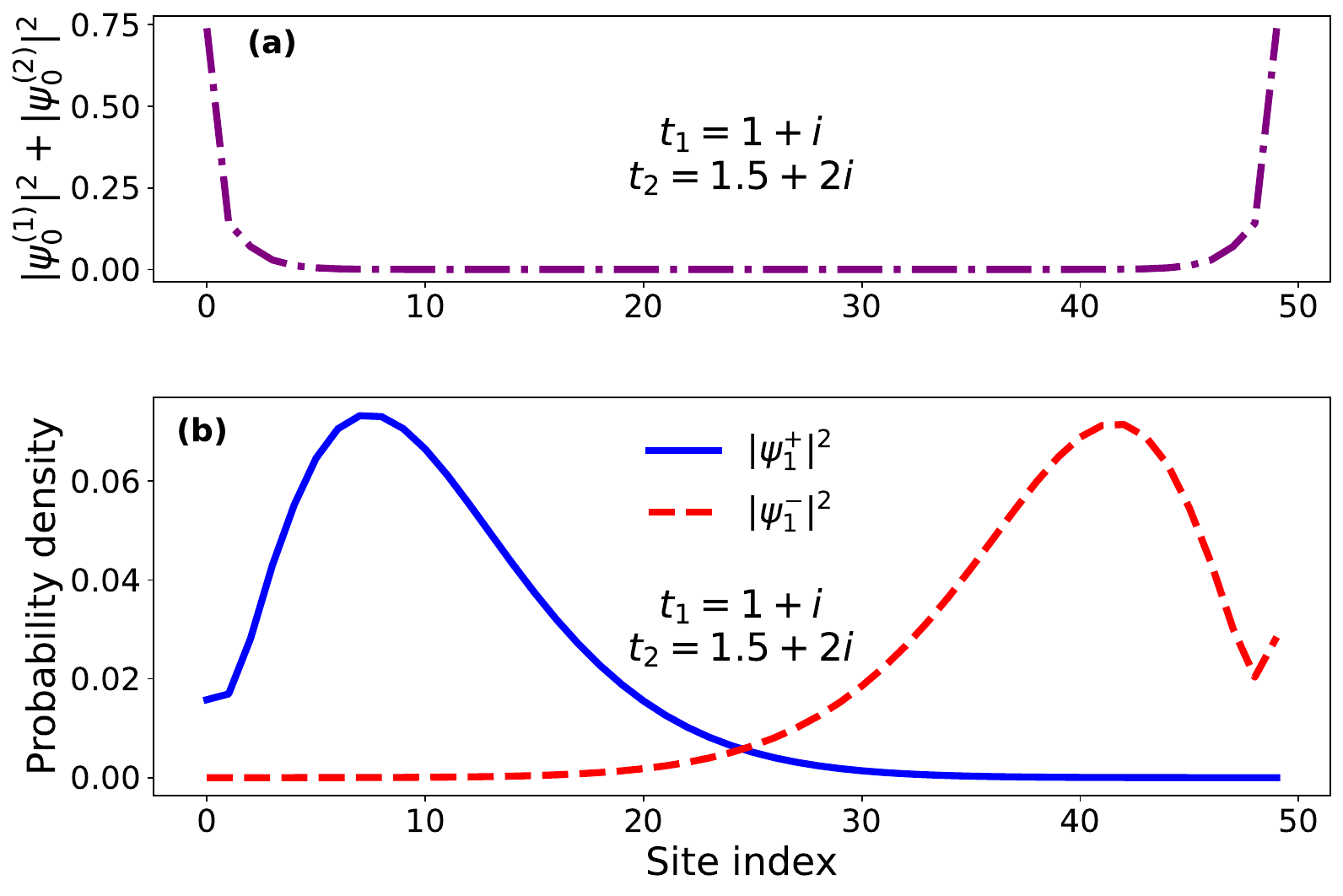}
    \caption{Probability density of zero modes and excited states of the non-Hermitian Kitaev chain with $L=50$ sites under open boundary conditions and $\mu=0.2+i,t_1=1+i,t_2=1.5+2i,\Delta_1=2+i,\Delta_2=1+3i$. \textbf{(a)} The total probability density of the two Majorana zero modes $\psi_0^{(1)}$ and $\psi_0^{(2)}$ is symmetric and localized at the edges. \textbf{(b)} $\psi_1^+$ and $\psi_1^-$ are the energy eigenstates corresponding to complex energies $\pm E, \text{Re}(E)>0$, where $|E|$ is the smallest absolute value of energy in the spectrum. They see an equal and opposite skin effect localization.}
    \label{fig:supp_fig1}
\end{figure}

\section{Symmetry analysis for non-Hermitian Kitaev chain} \label{appendix}

We now explicitly work out the discrete symmetries of the non-Hermitian Kitaev chain in different parameter regimes. The Hamiltonian for a finite chain is given as follows,
\begin{align}
       H^{\text{NH}} &= -\mu \sum_{j=1}^{L} c_j^{\dagger}c_j - \frac{1}{2} \sum_{j=1}^{L}(t_1 c_j^{\dagger}c_{j+1} + t_2 c_{j+1}^{\dagger}c_j + \Delta_1 e^{-i\varphi_{1}} c_j c_{j+1} + \Delta_2 e^{i\varphi_{2}} c_{j+1}^{\dagger}c_j^{\dagger} ) \,,
\end{align}
In the case of an infinite ($L\to\infty$) chain, the energy spectrum can be obtained by performing a Fourier transform i.e., $c_j = \frac{1}{\sqrt{L}}\sum_k e^{-ikj} c_k$ (for simplicity, the lattice constant is set to $1$), and obtain the Bogoliubov-de-Gennes (BdG) Hamiltonian, $H^{\text{NH}}_{\text{BdG}}=\frac{1}{2}\sum_k \mathcal{C}_k^{\dagger} H(k) \mathcal{C}_{k}$, where,
\begin{equation}
H(k)=\left(\begin{array}{cc}
-t_{1}e^{-ik}-t_{2}e^{ik}-2\mu & 2i\Delta_{2}e^{i\varphi_{2}}\sin k\\
-2i\Delta_{1}e^{-i\varphi_{1}}\sin k & t_{1}e^{ik}+t_{2}e^{-ik}+2\mu
\end{array}\right)
\end{equation}
and $\mathcal{C}_{k}^{\dagger} = \begin{pmatrix}
c_{k}^{\dagger} & c_{-k} \end{pmatrix}$. The discrete symmetries have been discussed in \cite{Kawabata_2022}. The spectral constraints arising from the possible symmetries of the system are summarized in the Table~\ref{tab:symmetry-and-spectrum-kspace}.
\begin{table}[b!]
    \centering
    \caption{Symmetries and their implications on the spectrum in momentum space}
    \label{tab:symmetry-and-spectrum-kspace}

    \renewcommand{\arraystretch}{1.3} 
    \setlength{\tabcolsep}{10pt}      

    \begin{tabular}{|l|c|}
        \hline
        \textbf{Symmetry} & \textbf{Eigenvalue pairs} \\
        \hline
        TRS: $T H^*(k) T^{-1} = H(-k)$ & $(E(k),\,E^*(-k))$ \\
        CS: $S H^\dagger(k) S^\dagger = -H(k)$ & $(E(k),\,-E^*(k))$ \\
        PHS: $C H^T(k) C^{-1} = -H(-k)$ & $(E(k),\,-E(-k))$ \\
        TRS$^\dagger$: $T H^T(k) T^{-1} = H(-k)$ & $(E(k),\,E(-k))$ \\
        CS$^\dagger$ (SLS): $\Gamma H(k) \Gamma^{-1} = -H(k)$ & $(E(k),\,-E(k))$ \\
        PHS$^\dagger$: $C H^*(k) C^{-1} = -H(-k)$ & $(E(k),\,-E^*(-k))$ \\
        \hline
    \end{tabular}
\end{table}

We now explore the possible symmteries of the non-Hermitian Kitaev chain and obtain Tabel~\ref{tab:sym_tab} in the main text. The symmetries are unitary relations between $2 \times 2$ matrices, and a necessary and sufficient check is given by Specht's theorem, which states that two $2 \times 2$ matrices $A$ and $B$ are unitarily equivalent if and only if the following 3 criteria are satisfied
\begin{equation}
    \Tr[A] = \Tr[B] \quad ; \quad \Tr[A^2] = \Tr[B^2] \quad ; \quad \Tr[AA^{\dagger}] = \Tr[BB^{\dagger}]
\end{equation}

\subsection{Time reversal symmetry (TRS)}

For time reversal symmetry, we need to check if $H(-k)$ and
$H(k)^{*}$are unitarily related. The matrix forms are given as follows,
\begin{align}
H(-k) & =\left(\begin{array}{cc}
-t_{1}e^{ik}-t_{2}e^{-ik}-2\mu & -2i\Delta_{2}e^{i\varphi_{2}}\sin k\\
2i\Delta_{1}e^{-i\varphi_{1}}\sin k & t_{1}e^{-ik}+t_{2}e^{ik}+2\mu
\end{array}\right)\\
H(k)^{*} & =\left(\begin{array}{cc}
-t_{1}^{*}e^{ik}-t_{2}^{*}e^{-ik}-2\mu^{*} & -2i\Delta_{2}e^{-i\varphi_{2}}\sin k\\
2i\Delta_{1}e^{i\varphi_{1}}\sin k & t_{1}^{*}e^{-ik}+t_{2}^{*}e^{ik}+2\mu^{*}
\end{array}\right)
\end{align}
and similarly, the Hermitian conjugates are,
\begin{align}
H^{\dagger}(-k) & =\left(\begin{array}{cc}
-t_{1}^{*}e^{-ik}-t_{2}^{*}e^{ik}-2\mu^{*} & -2i\Delta_{1}e^{i\varphi_{1}}\sin k\\
2i\Delta_{2}e^{-i\varphi_{2}}\sin k & t_{1}^{*}e^{ik}+t_{2}^{*}e^{-ik}+2\mu^{*}
\end{array}\right)\\
H^{\dagger}(k)^{*} & =\left(\begin{array}{cc}
-t_{1}e^{-ik}-t_{2}e^{ik}-2\mu & -2i\Delta_{1}e^{-i\varphi_{1}}\sin k\\
2i\Delta_{2}e^{i\varphi_{2}}\sin k & t_{1}e^{ik}+t_{2}e^{-ik}+2\mu^{*}
\end{array}\right)
\end{align}
From here we evaluate the following,
\begin{align}
\text{Tr}\big[H(-k)\big] & =(t_{2}-t_{1})e^{ik}+(t_{1}-t_{2})e^{-ik}\\
\text{Tr}\big[H(k)^{*}\big] & =(t_{2}^{*}-t_{1}^{*})e^{ik}+(t_{1}^{*}-t_{2}^{*})e^{-ik}
\end{align}
and similarly,
\begin{align}
\text{Tr}\big[H(-k)^{2}\big]  
 & =(t_{1}^{2}+t_{2}^{2})e^{i2k}+(t_{1}^{2}+t_{2}^{2})e^{-i2k}+4\mu(t_{1}+t_{2})e^{ik}+4\mu(t_{1}+t_{2})e^{-ik}+8\mu^{2}\notag \\
 & \quad+4t_{1}t_{2}+8\Delta_{1}\Delta_{2}e^{i\varphi_{2}-i\varphi_{1}}\sin^{2}k\\
\text{Tr}\big[H(k)^{*2}\big]
 & =(t_{1}^{*2}+t_{2}^{*2})e^{-i2k}+(t_{1}^{*2}+t_{2}^{*2})e^{i2k}+4\mu^{*}(t_{1}^{*}+t_{2}^{*})e^{-ik}+4\mu^{*}(t_{1}^{*}+t_{2}^{*})e^{ik}+8\mu^{*2}\notag \\
 & \quad+4t_{1}^{*}t_{2}^{*}+8\Delta_{1}\Delta_{2}e^{i\varphi_{1}-i\varphi_{2}}\sin^{2}k
\end{align}
Finally, we evaluate,
\begin{align}
\text{Tr}\big[H(k)^{*}H^{\dagger}(k)^{*}\big] 
 & =|t_{1}|^{2}+t_{1}^{*}t_{2}e^{i2k}+2\mu t_{1}^{*}e^{ik}+t_{1}t_{2}^{*}e^{-i2k}+|t_{2}|^{2}+2\mu t_{2}^{*}e^{-ik}+2\mu^{*}t_{1}e^{-ik}+2\mu^{*}t_{2}e^{ik}\notag \\
 & \quad+|t_{1}|^{2}+t_{1}^{*}t_{2}e^{-i2k}+2\mu^{*}t_{1}^{*}e^{-ik}+t_{1}t_{2}^{*}e^{i2k}+|t_{2}|^{2}+2\mu^{*}t_{2}^{*}e^{ik}+2\mu^{*}t_{1}e^{ik}+2\mu^{*}t_{2}e^{-ik}\notag \\
 & \quad+4\Delta_{2}^{2}\sin^{2}k+4\Delta_{1}^{2}\sin^{2}k + 8|\mu|^{2}\\
\text{Tr}\big[H(-k)H^{\dagger}(-k)\big] 
 & =|t_{1}|^{2}+t_{1}t_{2}^{*}e^{i2k}+2\mu^{*}t_{1}e^{ik}+t_{1}^{*}t_{2}e^{-2ik}+|t_{2}|^{2}+2\mu^{*}t_{2}e^{-ik}+2\mu t_{1}^{*}e^{-ik}+2\mu t_{2}^{*}e^{ik}\notag \\
 & \quad+|t_{1}|^{2}+t_{1}t_{2}^{*}e^{-i2k}+2\mu^{*}t_{1}e^{-ik}+t_{1}^{*}t_{2}e^{2ik}+|t_{2}|^{2}+2\mu^{*}t_{2}e^{ik}+2\mu t_{1}^{*}e^{ik}+2\mu t_{2}^{*}e^{-ik}\notag \\
 & \quad+4\Delta_{1}^{2}\sin^{2}k+4\Delta_{2}^{2}\sin^{2}k+8|\mu|^{2}
\end{align}
Hence, TRS exists if,
\begin{align}
t_{1},\,t_{2} & \in\mathbb{R}\notag \\
\varphi_{1} & =\varphi_{2}\\
\mu & \in\mathbb{R}\notag 
\end{align}

\subsection{Particle hole symmetry (PHS)}

For particle hole symmetry (PHS), there must exist a unitary relation between
$-H(-k)$ and $H(k)^{T}$. The full matrix forms are given as follows,
\begin{align}
-H(-k) & =\left(\begin{array}{cc}
t_{1}e^{ik}+t_{2}e^{-ik}+2\mu & 2i\Delta_{2}e^{i\varphi_{2}}\sin k\\
-2i\Delta_{1}e^{-i\varphi_{1}}\sin k & -t_{1}e^{-ik}-t_{2}e^{ik}-2\mu
\end{array}\right)\\
H(k)^{T} & =\left(\begin{array}{cc}
-t_{1}e^{-ik}-t_{2}e^{ik}-2\mu & -2i\Delta_{1}e^{-i\varphi_{1}}\sin k\\
2i\Delta_{2}e^{i\varphi_{2}}\sin k & t_{1}e^{ik}+t_{2}e^{-ik}+2\mu
\end{array}\right)
\end{align}
From here, we evaluate the Hermitian conjugates as follows,
\begin{align}
-H^{\dagger}(-k) & =\left(\begin{array}{cc}
t_{1}^{*}e^{-ik}+t_{2}^{*}e^{ik}+2\mu^{*} & 2i\Delta_{1}e^{i\varphi_{1}}\sin k\\
-2i\Delta_{2}e^{-i\varphi_{2}}\sin k & -t_{1}^{*}e^{ik}-t_{2}^{*}e^{-ik}-2\mu^{*}
\end{array}\right)\\
H^{\dagger}(k)^{T} & =\left(\begin{array}{cc}
-t_{1}^{*}e^{ik}-t_{2}^{*}e^{-ik}-2\mu^{*} & -2i\Delta_{2}e^{-i\varphi_{2}}\sin k\\
2i\Delta_{1}e^{i\varphi_{1}}\sin k & t_{1}^{*}e^{-ik}+t_{2}^{*}e^{ik}+2\mu^{*}
\end{array}\right)
\end{align}
Using this, we evaluate the following,
\begin{align}
\text{Tr}\big[-H(-k)\big] & =(t_{1}-t_{2})e^{ik}+(t_{2}-t_{1})e^{-ik}\\
\text{Tr}\big[H(k)^{T}\big] & =(t_{1}-t_{2})e^{ik}+(t_{2}-t_{1})e^{-ik}
\end{align}
and 
\begin{align}
\text{Tr}\big[H(-k)^{2}\big] & =(t_{1}^{2}+t_{2}^{2})e^{i2k}+(t_{1}^{2}+t_{2}^{2})e^{-i2k}+4\mu(t_{1}+t_{2})e^{ik}+4\mu(t_{1}+t_{2})e^{-ik}\notag \\
&+8\mu^{2}+4t_{1}t_{2}+8\Delta_{1}\Delta_{2}e^{i\varphi_{2}-i\varphi_{1}}\sin^{2}k\\
\text{Tr}\Big(\big[H(k)^{T}\big]^{2}\Big) & =(t_{1}^{2}+t_{2}^{2})e^{-i2k}+(t_{1}^{2}+t_{2}^{2})e^{i2k}+4\mu(t_{1}+t_{2})e^{-ik}+4\mu(t_{1}+t_{2})e^{ik}\notag \\
&+8\mu^{2}+4t_{1}t_{2}+8\Delta_{1}\Delta_{2}e^{i\varphi_{2}-i\varphi_{1}}\sin^{2}k
\end{align}
Finally, we evaluate,
\begin{align}
\text{Tr}\big[H(-k)H^{\dagger}(-k)\big] 
 & =|t_{1}|^{2}+t_{1}t_{2}^{*}e^{2ik}+2\mu^{*}t_{1}e^{ik}+t_{1}^{*}t_{2}e^{-2ik}+|t_{2}|^{2}+2\mu^{*}t_{2}e^{-ik}+2\mu t_{1}^{*}e^{-ik}+2\mu t_{2}^{*}e^{ik}\notag \\
 & \quad+|t_{1}|^{2}+t_{1}t_{2}^{*}e^{-2ik}+2\mu^{*}t_{1}e^{-ik}+t_{1}^{*}t_{2}e^{2ik}+|t_{2}|^{2}+2\mu^{*}t_{2}e^{ik}+2\mu t_{1}^{*}e^{ik}+2\mu t_{2}^{*}e^{-ik}\notag \\
 & \quad+4\Delta_{1}^{2}\sin^{2}k+4\Delta_{2}^{2}\sin^{2}k + 8|\mu|^{2}\\
\text{Tr}\big[H(k)^{T}H^{\dagger}(k)^{T}\big] 
 & =|t_{1}|^{2}+t_{1}t_{2}^{*}e^{-2ik}+2\mu^{*}t_{1}e^{-ik}+t_{1}^{*}t_{2}e^{2ik}+|t_{2}|^{2}+2\mu^{*}t_{2}e^{ik}+2\mu t_{1}^{*}e^{ik}+2\mu t_{2}^{*}e^{-ik}\notag \\
 & \quad+|t_{1}|^{2}+t_{1}t_{2}^{*}e^{2ik}+2\mu^{*}t_{1}e^{ik}+t_{1}^{*}t_{2}e^{-2ik}+|t_{2}|^{2}+2\mu^{*}t_{2}e^{-ik}+2\mu t_{1}^{*}e^{-ik}+2\mu t_{2}^{*}e^{ik}\notag \\
 & \quad+4\Delta_{1}^{2}\sin^{2}k+4\Delta_{2}^{2}\sin^{2}k+8|\mu|^{2}
\end{align}
From here, we find that $-H(-k)$ and $H(k)^{T}$ are always unitarily related, i.e., PHS holds for all values of $t_{1},\,t_{2},\,\mu,\,\varphi_{1},\,\varphi_{2}$.

The generator of PHS for the non-Hermitian Kitaev chain is $\sigma_x$, i.e. we have,
\begin{equation}
    \sigma_x H(k)^T \sigma_x = -H(-k).
\end{equation}
In the most general parameter regime, PHS is the only symmetry of the non-Hermitian Kitaev chain, Table~\ref{tab:sym_tab}. Following the symmetry-based topological classification of \cite{PhysRevX.9.041015}, we observe that the generator of PHS squares to identity, i.e. $\sigma_x^2 = \mathbbm{1}$. Using this and the fact that the spectrum has a line gap, we conclude that the non-Hermitian Kitaev chain lies in class D, which has a $\mathbb{Z}_2$ topological invariant. 

\subsection{Adjoint time reversal symmetry (TRS$^{\dagger}$)}

For adjoint time reversal symmetry (TRS$^{\dagger}$), $H(-k)$ and
$H(k)^{T}$ are unitarily related. The full matrix forms are,
\begin{align}
H(-k) & =\left(\begin{array}{cc}
-t_{1}e^{ik}-t_{2}e^{-ik}-2\mu & -2i\Delta_{2}e^{i\varphi_{2}}\sin k\\
2i\Delta_{1}e^{-i\varphi_{1}}\sin k & t_{1}e^{-ik}+t_{2}e^{ik}+2\mu
\end{array}\right)\\
H(k)^{T} & =\left(\begin{array}{cc}
-t_{1}e^{-ik}-t_{2}e^{ik}-2\mu & -2i\Delta_{1}e^{-i\varphi_{1}}\sin k\\
2i\Delta_{2}e^{i\varphi_{2}}\sin k & t_{1}e^{ik}+t_{2}e^{-ik}+2\mu
\end{array}\right)
\end{align}
and the Hermtian conjugates are,
\begin{align}
H^{\dagger}(-k) & =\left(\begin{array}{cc}
-t_{1}^{*}e^{-ik}-t_{2}^{*}e^{ik}-2\mu^{*} & -2i\Delta_{1}e^{i\varphi_{1}}\sin k\\
2i\Delta_{2}e^{-i\varphi_{2}}\sin k & t_{1}^{*}e^{ik}+t_{2}^{*}e^{-ik}+2\mu^{*}
\end{array}\right)\\
H^{\dagger}(k)^{T} & =\left(\begin{array}{cc}
-t_{1}^{*}e^{ik}-t_{2}^{*}e^{-ik}-2\mu^{*} & -2i\Delta_{2}e^{-i\varphi_{2}}\sin k\\
2i\Delta_{1}e^{i\varphi_{1}}\sin k & t_{1}^{*}e^{-ik}+t_{2}^{*}e^{ik}+2\mu^{*}
\end{array}\right)
\end{align}
Next, we evaluate the following,
\begin{align}
\text{Tr}\big[H(-k)\big] & =(t_{2}-t_{1})e^{ik}+(t_{1}-t_{2})e^{-ik}\\
\text{Tr}\big[H(k)^{T}\big] & =(t_{1}-t_{2})e^{ik}+(t_{2}-t_{1})e^{-ik}
\end{align}
similarly,
\begin{align}
\text{Tr}\big[H(-k)^{2}\big] & =(t_{1}^{2}+t_{2}^{2})e^{i2k}+(t_{1}^{2}+t_{2}^{2})e^{-i2k}+4\mu(t_{1}+t_{2})e^{ik}+4\mu(t_{1}+t_{2})e^{-ik}\notag \\
&+8\mu^{2}+4t_{1}t_{2}+8\Delta_{1}\Delta_{2}e^{i\varphi_{2}-i\varphi_{1}}\sin^{2}k\\
\text{Tr}\Big(\big[H(k)^{T}\big]^{2}\Big) & =(t_{1}^{2}+t_{2}^{2})e^{-i2k}+(t_{1}^{2}+t_{2}^{2})e^{i2k}+4\mu(t_{1}+t_{2})e^{-ik}+4\mu(t_{1}+t_{2})e^{ik}\notag \\
&+8\mu^{2}+4t_{1}t_{2}+8\Delta_{1}\Delta_{2}e^{i\varphi_{2}-i\varphi_{1}}\sin^{2}k
\end{align}
Finally, we evaluate,
\begin{align}
\text{Tr}\big[H(-k)H^{\dagger}(-k)\big]
 & =|t_{1}|^{2}+t_{1}t_{2}^{*}e^{i2k}+2\mu^{*}t_{1}e^{ik}+t_{1}^{*}t_{2}e^{-i2k}+|t_{2}|^{2}+2\mu^{*}t_{2}e^{-ik}+2\mu t_{1}^{*}e^{-ik}+2\mu t_{2}e^{ik}\notag \\
 & \quad+|t_{1}|^{2}+t_{1}t_{2}^{*}e^{-i2k}+2\mu^{*}t_{1}e^{-ik}+t_{1}^{*}t_{2}e^{i2k}+|t_{2}|^{2}+2\mu^{*}t_{2}e^{ik}+2\mu t_{1}^{*}e^{ik}+2\mu t_{2}e^{-ik}\notag \\
 & \quad+4\Delta_{1}^{2}\sin^{2}k+4\Delta_{2}^{2}\sin^{2}k + 8|\mu|^{2}\\
\text{Tr}\big[H(k)^{T}H^{\dagger}(k)^{T}\big] & =|t_{1}|^{2}+t_{1}t_{2}^{*}e^{-2ik}+2\mu^{*}t_{1}e^{-ik}+t_{1}^{*}t_{2}e^{2ik}+|t_{2}|^{2}+2\mu^{*}t_{2}e^{ik}+2\mu t_{1}^{*}e^{ik}+2\mu t_{2}^{*}e^{-ik}\notag \\
 & \quad+|t_{1}|^{2}+t_{1}t_{2}^{*}e^{2ik}+2\mu^{*}t_{1}e^{ik}+t_{1}^{*}t_{2}e^{-2ik}+|t_{2}|^{2}+2\mu^{*}t_{2}e^{-ik}+2\mu t_{1}^{*}e^{-ik}+2\mu t_{2}^{*}e^{ik}\notag \\
 & \quad+4\Delta_{1}^{2}\sin^{2}k+4\Delta_{2}^{2}\sin^{2}k +8|\mu|^{2}
\end{align}
Hence, we find the following conditions for TRS$^{\dagger}$ to hold,
\begin{align}
\mu & \in\mathbb{C}\notag \\
t_{1} & =t_{2}\in\mathbb{C}\\
\forall \,  & \varphi_{1},\,\varphi_{2}\notag 
\end{align}

\subsection{Adjoint particle hole symmetry (PHS$^{\dagger}$)}

For adjoint particle hole symmetry (PHS$^{\dagger}$), there exists
a unitary relation between $-H(-k)$ and $H(k)^{*}$. The matrix
forms are,
\begin{align}
-H(-k) & =\left(\begin{array}{cc}
t_{1}e^{ik}+t_{2}e^{-ik}+2\mu & 2i\Delta_{2}e^{i\varphi_{2}}\sin k\\
-2i\Delta_{1}e^{-i\varphi_{1}}\sin k & -t_{1}e^{-ik}-t_{2}e^{ik}-2\mu
\end{array}\right)\\
H(k)^{*} & =\left(\begin{array}{cc}
-t_{1}^{*}e^{ik}-t_{2}^{*}e^{-ik}-2\mu^{*} & -2i\Delta_{2}e^{-i\varphi_{2}}\sin k\\
2i\Delta_{1}e^{i\varphi_{1}}\sin k & t_{1}^{*}e^{-ik}+t_{2}^{*}e^{ik}+2\mu^{*}
\end{array}\right)
\end{align}
From here, we obtain the Hermitian conjugates,
\begin{align}
-H^{\dagger}(-k) & =\left(\begin{array}{cc}
t_{1}^{*}e^{-ik}+t_{2}^{*}e^{ik}+2\mu^{*} & 2i\Delta_{1}e^{i\varphi_{1}}\sin k\\
-2i\Delta_{2}e^{-i\varphi_{2}}\sin k & -t_{1}^{*}e^{ik}-t_{2}^{*}e^{-ik}-2\mu^{*}
\end{array}\right)\\
H^{\dagger}(k)^{*} & =\left(\begin{array}{cc}
-t_{1}e^{-ik}-t_{2}e^{ik}-2\mu & -2i\Delta_{1}e^{-i\varphi_{1}}\sin k\\
2i\Delta_{2}e^{i\varphi_{2}}\sin k & t_{1}e^{ik}+t_{2}e^{-ik}+2\mu^{*}
\end{array}\right)
\end{align}
We now have,
\begin{align}
\text{Tr}\big[-H(-k)\big] & =(t_{1}-t_{2})e^{ik}+(t_{2}-t_{1})e^{-ik}\\
\text{Tr}\big[H(k)^{*}\big] & =(t_{2}^{*}-t_{1}^{*})e^{ik}+(t_{1}^{*}-t_{2}^{*})e^{-ik}
\end{align}
Similarly, we obtain,
\begin{align}
\text{Tr}\big[H(-k)^{2}\big] & =(t_{1}^{2}+t_{2}^{2})e^{i2k}+(t_{1}^{2}+t_{2}^{2})e^{-i2k}+4\mu(t_{1}+t_{2})e^{ik}+4\mu(t_{1}+t_{2})e^{-ik}+8\mu^{2}\notag \\
 & \quad+4t_{1}t_{2}+8\Delta_{1}\Delta_{2}e^{i\varphi_{2}-i\varphi_{1}}\sin^{2}k\\
\text{Tr}\big[H(k)^{*2}\big] & =(t_{1}^{*2}+t_{2}^{*2})e^{-i2k}+(t_{1}^{*2}+t_{2}^{*2})e^{i2k}+4\mu^{*}(t_{1}^{*}+t_{2}^{*})e^{-ik}+4\mu^{*}(t_{1}^{*}+t_{2}^{*})e^{ik}+8\mu^{*2}\notag \\
 & \quad+4t_{1}^{*}t_{2}^{*}+8\Delta_{1}\Delta_{2}e^{i\varphi_{1}-i\varphi_{2}}\sin^{2}k
\end{align}
Finally, we have,
\begin{align}
\text{Tr}\big[H(-k)H^{\dagger}(-k)\big] & =|t_{1}|^{2}+t_{1}t_{2}^{*}e^{2ik}+2\mu^{*}t_{1}e^{ik}+t_{1}^{*}t_{2}e^{-2ik}+|t_{2}|^{2}+2\mu^{*}t_{2}e^{-ik}+2\mu t_{1}^{*}e^{-ik}+2\mu t_{2}^{*}e^{ik}\notag \\
 & \quad+|t_{1}|^{2}+t_{1}t_{2}^{*}e^{-2ik}+2\mu^{*}t_{1}e^{-ik}+t_{1}^{*}t_{2}e^{2ik}+|t_{2}|^{2}+2\mu^{*}t_{2}e^{ik}+2\mu t_{1}^{*}e^{ik}+2\mu t_{2}^{*}e^{-ik}\notag \\
 & \quad+4\Delta_{1}^{2}\sin^{2}k+4\Delta_{2}^{2}\sin^{2}k+8|\mu|^{2}\\
\text{Tr}\big[H(k)^{*}H^{\dagger}(k)^{*}\big] & =|t_{1}|^{2}+t_{1}^{*}t_{2}e^{i2k}+2\mu t_{1}^{*}e^{ik}+t_{1}t_{2}^{*}e^{-i2k}+|t_{2}|^{2}+2\mu t_{2}^{*}e^{-ik}+2\mu^{*}t_{1}e^{-ik}+2\mu^{*}t_{2}e^{ik}\notag \\
 & \quad+|t_{1}|^{2}+t_{1}^{*}t_{2}e^{-i2k}+2\mu^{*}t_{1}^{*}e^{-ik}+t_{1}t_{2}^{*}e^{i2k}+|t_{2}|^{2}+2\mu^{*}t_{2}^{*}e^{ik}+2\mu^{*}t_{1}e^{ik}+2\mu^{*}t_{2}e^{-ik}\notag \\
 & \quad+4\Delta_{2}^{2}\sin^{2}k+4\Delta_{1}^{2}\sin^{2}k+8|\mu|^{2}
\end{align}
Hence, $-H(-k)$ and $H(k)^{*}$ are unitarily equivalent and PHS$^{\dagger}$ holds if we have,
\begin{align}
t_{1} & =t_{2}^{*}\in\mathbb{C}\notag \\
\mu & \in\mathbb{R}\\
\varphi_{1} & =\varphi_{2}\notag 
\end{align}

\subsection{Chiral symmetry (CS)}

For the system to have chiral symmetry (CS), the Hamiltonians $-H(k)$
and $H(k)^{\dagger}$ must be unitarily related. The explicit matrix forms
are given as follows,
\begin{align}
-H(k) & =\left(\begin{array}{cc}
t_{1}e^{-ik}+t_{2}e^{ik}+2\mu & -2i\Delta_{2}e^{i\varphi_{2}}\sin k\\
2i\Delta_{1}e^{-i\varphi_{1}}\sin k & -t_{1}e^{ik}-t_{2}e^{-ik}-2\mu
\end{array}\right)\\
H(k)^{\dagger} & =\left(\begin{array}{cc}
-t_{1}^{*}e^{ik}-t_{2}^{*}e^{-ik}-2\mu^{*} & 2i\Delta_{1}e^{i\varphi_{1}}\sin k\\
-2i\Delta_{2}e^{-i\varphi_{2}}\sin k & t_{1}^{*}e^{-ik}+t_{2}^{*}e^{ik}+2\mu^{*}
\end{array}\right)
\end{align}
The Hermitian conjugates are given as follows,
\begin{align}
-H(k)^{\dagger} & =\left(\begin{array}{cc}
t_{1}^{*}e^{ik}+t_{2}^{*}e^{-ik}+2\mu^{*} & -2i\Delta_{1}e^{i\varphi_{1}}\sin k\\
2i\Delta_{1}e^{-i\varphi_{2}}\sin k & -t_{1}^{*}e^{-ik}-t_{2}^{*}e^{ik}-2\mu^{*}
\end{array}\right)\\
H(k) & =\left(\begin{array}{cc}
-t_{1}e^{-ik}-t_{2}e^{ik}-2\mu & 2i\Delta_{2}e^{i\varphi_{2}}\sin k\\
-2i\Delta_{1}e^{-i\varphi_{1}}\sin k & t_{1}e^{ik}+t_{2}e^{-ik}+2\mu
\end{array}\right)
\end{align}
From here we obtain the following,
\begin{align}
\text{Tr}\big[-H(k)\big] & =(t_{1}-t_{2})e^{-ik}+(t_{2}-t_{1})e^{ik}\\
\text{Tr}\big[H(k)^{\dagger}\big] & =(t_{1}^{*}-t_{2}^{*})e^{-ik}+(t_{2}^{*}-t_{1}^{*})e^{ik}
\end{align}
Similarly, we have,
\begin{align}
\text{Tr}\big[H(k)^{2}\big] & =(t_{1}^{2}+t_{2}^{2})e^{i2k}+(t_{1}^{2}+t_{2}^{2})e^{-i2k}+4\mu(t_{1}+t_{2})e^{ik}+4\mu(t_{1}+t_{2})e^{-ik}\\
&+4\mu^{2}+4t_{1}t_{2}+8\Delta_{1}\Delta_{2}e^{i\varphi_{2}-i\varphi_{1}}\sin^{2}k\\
\text{Tr}\Big(\big[H(k)^{\dagger}\big]^{2}\Big) & =(t_{1}^{*2}+t_{2}^{*2})e^{i2k}+(t_{1}^{*2}+t_{2}^{*2})e^{-i2k}+4\mu(t_{1}^{*}+t_{2}^{*})e^{ik}+4\mu(t_{1}^{*}+t_{2}^{*})e^{-ik}\\
&+4\mu^{2}+4t_{1}^{*}t_{2}^{*}+8\Delta_{1}\Delta_{2}e^{i\varphi_{1}-i\varphi_{2}}\sin^{2}k
\end{align}
And finally,
\begin{align}
\text{Tr}\big[H(k)H^{\dagger}(k)\big] 
 & =|t_{1}|^{2}+t_{1}t_{2}^{*}e^{-i2k}+2\mu^{*}t_{1}e^{-ik}+t_{1}^{*}t_{2}e^{i2k}+|t_{2}|^{2}+2\mu^{*}t_{2}e^{ik}+2\mu t_{1}^{*}e^{ik}+2\mu t_{2}^{*}e^{-ik}\notag \\
 & \quad+|t_{1}|^{2}+t_{1}t_{2}^{*}e^{i2k}+2\mu^{*}t_{1}e^{ik}+t_{1}^{*}t_{2}e^{-i2k}+|t_{2}|^{2}+2\mu^{*}t_{2}e^{-ik}+2\mu t_{1}^{*}e^{-ik}+2\mu t_{2}^{*}e^{ik}\notag \\
 & \quad+4\Delta_{1}^{2}\sin^{2}k+4\Delta_{2}^{2}\sin^{2}k+8|\mu|^{2}\\
\text{Tr}\big[H(k)^{\dagger}H(k)\big] 
 & =|t_{1}|^{2}+t_{1}^{*}t_{2}e^{i2k}+2\mu t_{1}^{*}e^{ik}+t_{1}t_{2}^{*}e^{-i2k}+|t_{2}|^{2}+2\mu t_{2}^{*}e^{-ik}+2\mu^{*}t_{1}e^{-ik}+2\mu^{*}t_{2}e^{ik}\notag \\
 & \quad+|t_{1}|^{2}+t_{1}^{*}t_{2}e^{-i2k}+2\mu t_{1}^{*}e^{-ik}+t_{1}t_{2}^{*}e^{i2k}+|t_{2}|^{2}+2\mu t_{2}^{*}e^{ik}+2\mu^{*}t_{1}e^{ik}+2\mu^{*}t_{2}e^{-ik}\notag \\
 & \quad+4\Delta_{1}^{2}\sin^{2}k+4\Delta_{2}^{2}\sin^{2}k+8|\mu|^{2}
\end{align}
Hence, unitary equivalence between $-H(k)$ and $H(k)^{\dagger}$ exists and CS holds
if we have,
\begin{align}
t_{1},\,t_{2},\,\mu & \in\mathbb{R}\notag \\
\varphi_{1} & =\varphi_{2}
\end{align}

\subsection{Adjoint chiral symmetry (CS$^{\dagger}$)}

For the system to have adjoint chiral symmetry (CS$^{\dagger}$),
the Hamiltonians $-H(k)$ and $H(k)$ are unitarily related. The explicit
matrix forms are given as follows,
\begin{align}
-H(k) & =\left(\begin{array}{cc}
t_{1}e^{-ik}+t_{2}e^{ik}+2\mu & -2i\Delta_{2}e^{i\varphi_{2}}\sin k\\
2i\Delta_{1}e^{-i\varphi_{1}}\sin k & -t_{1}e^{ik}-t_{2}e^{-ik}-2\mu
\end{array}\right)\\
H(k) & =\left(\begin{array}{cc}
-t_{1}e^{-ik}-t_{2}e^{ik}-2\mu & 2i\Delta_{2}e^{i\varphi_{2}}\sin k\\
-2i\Delta_{1}e^{-i\varphi_{1}}\sin k & t_{1}e^{ik}+t_{2}e^{-ik}+2\mu
\end{array}\right)
\end{align}
The Hermitian conjugates are given as follows,
\begin{align}
-H(k)^{\dagger} & =\left(\begin{array}{cc}
t_{1}^{*}e^{ik}+t_{2}^{*}e^{-ik}+2\mu^{*} & -2i\Delta_{1}e^{i\varphi_{1}}\sin k\\
2i\Delta_{1}e^{-i\varphi_{2}}\sin k & -t_{1}^{*}e^{-ik}-t_{2}^{*}e^{ik}-2\mu^{*}
\end{array}\right)\\
H(k)^{\dagger} & =\left(\begin{array}{cc}
-t_{1}^{*}e^{ik}-t_{2}^{*}e^{-ik}-2\mu^{*} & 2i\Delta_{1}e^{i\varphi_{1}}\sin k\\
-2i\Delta_{2}e^{-i\varphi_{2}}\sin k & t_{1}^{*}e^{-ik}+t_{2}^{*}e^{ik}+2\mu^{*}
\end{array}\right)
\end{align}
From here we obtain the following,
\begin{align}
\text{Tr}\big[-H(k)\big] & =(t_{1}-t_{2})e^{-ik}+(t_{2}-t_{1})e^{ik}\\
\text{Tr}\big[H(k)\big] & =(t_{2}-t_{1})e^{-ik}+(t_{1}-t_{2})e^{ik}
\end{align}
Since our matrices are related by an overall sign, we have,
\begin{align}
\text{Tr}\big[H(k)^{2}\big] & =\text{Tr}\big[\{-H(k)\}^{2}\big]\\
\text{Tr}\big[H(k)H(k)^{\dagger}\big] & =\text{Tr}\big[\{-H(k)\}\{-H^{\dagger}(k)\}\big]
\end{align}
Hence, we have CS$^{\dagger}$ if,
\begin{align}
t_{1} & =t_{2}\in\mathbb{C}\notag \\
\forall & \varphi_{1},\,\varphi_{2},\,\mu
\end{align}
\end{widetext}
\end{appendix}

\bibliography{references}

\end{document}